\documentclass[sts]{imsart}

\startlocaldefs

\usepackage[figuresright]{rotating}
\usepackage{tabu}
\usepackage{hyperref}
\usepackage{amsmath}
\usepackage{amssymb}
\usepackage{amsthm}
\usepackage{mathtools}
\usepackage{pifont}
\usepackage{lineno}
\usepackage{graphicx}
\usepackage{enumerate}
\usepackage{rotating}
\usepackage[round]{natbib}
\usepackage{scrextend}
\usepackage{datetime}
\usepackage{array}
\usepackage{multicol}
\usepackage{subcaption}
\usepackage{chngcntr}
\usepackage{pbox}
\usepackage{pgf}
\usepackage{tikz}
\usetikzlibrary{arrows,automata}
\usetikzlibrary{positioning}
\usetikzlibrary{shapes}
\usetikzlibrary{decorations.text}

\tikzset{
    state/.style={
           rectangle,
           rounded corners,
           draw=black, very thick,
           minimum height=2em,
           inner sep=2pt,
           text centered,
           },
}

\definecolor{coolgrey}{rgb}{0.55, 0.57, 0.67}
\definecolor{lightgray}{rgb}{0.78, 0.78, 0.78}

\newcommand{\s}{\mathbf{s}}
\newcommand{\h}{\mathbf{h}}

\newcommand{\A}{\mathbf{A}}
\newcommand{\G}{\mathbf{G}}
\newcommand{\D}{\mathcal{D}}
\newcommand{\R}{\mathbb{R}}
\newcommand{\Z}{\mathbb{Z}}

\newcommand{\bg}{\boldsymbol{\gamma}}
\newcommand{\bo}{\boldsymbol{\omega}}

\newcommand{\bL}{\mathbf{\Lambda}}
\newtheorem{definition}[subsection]{Definition}

\numberwithin{equation}{section}

\defcitealias{lu2005test}{LZ}
\defcitealias{guan2004isotropy}{GSC}
\defcitealias{maity2012test}{MS}

\endlocaldefs

\begin{document}

\begin{frontmatter}

\title{A Review of Nonparametric Hypothesis Tests of Isotropy Properties in Spatial Data}
\runtitle{Nonparametric Hypothesis Tests of Isotropy}


\author{\fnms{Zachary D.} \snm{Weller}\ead[label=e1]{wellerz@stat.colostate.edu}}\address{\printead{e1}}\and \author{\fnms{Jennifer A.} \snm{Hoeting}\ead[label=e2]{jah@stat.colostate.edu}} \address{\printead{e2}}

\affiliation{Department of Statistics, Colorado State University}

\thankstext{t1}{Weller's work was supported by the National Science Foundation Research Network on Statistics in the Atmospheric and Ocean Sciences (STATMOS) (DMS-1106862). Hoeting's research was supported by the National Science Foundation (AGS-1419558).}
\thankstext{t2}{The authors would like to thank Peter Guttorp and Alexandra Schmidt for organizing the Pan-American Advanced Study Institute (PASI) on spatiotemporal statistics in June 2014 which inspired this work.}

\corref{Zachary D. Weller}
\runauthor{Zachary D. Weller}

\begin{abstract}
An important aspect of modeling spatially-referenced data is appropriately specifying the covariance function of the random field. A practitioner working with spatial data is presented a number of choices regarding the structure of the dependence between observations. One of these choices is determining whether or not an isotropic covariance function is appropriate. Isotropy implies that spatial dependence does not depend on the direction of the spatial separation between sampling locations. Misspecification of isotropy properties (directional dependence) can lead to misleading inferences, e.g., inaccurate predictions and parameter estimates. A researcher may use graphical diagnostics, such as directional sample variograms, to decide whether the assumption of isotropy is reasonable. These graphical techniques can be difficult to assess, open to subjective interpretations, and misleading. Hypothesis tests of the assumption of isotropy may be more desirable. To this end, a number of tests of directional dependence have been developed using both the spatial and spectral representations of random fields. We provide an overview of nonparametric methods available to test the hypotheses of isotropy and symmetry in spatial data. We summarize test properties, discuss important considerations and recommendations in choosing and implementing a test, compare several of the methods via a simulation study, and propose a number of open research questions. Several of the reviewed methods can be implemented in \texttt{R} using our package \texttt{spTest}, available on \texttt{CRAN}.
\end{abstract}



\end{frontmatter}

\section{Introduction}\label{intro}

Early spatial models relied on the simplifying assumptions that the covariance function is stationary and isotropic. With the emergence of new sources of spatial data, for instance, remote sensing via satellite, climate model output, or environmental monitoring, a variety of methods and models have been developed that relax these assumptions. In the case of anisotropy, there are a number of methods for modeling both zonal anisotropy (\citeauthor{journel1978mining}, \citeyear{journel1978mining}, pg.~179-184; \citeauthor{ecker2003spatial}, \citeyear{ecker2003spatial}; \citeauthor{schab}, \citeyear{schab}, pg.~152; \citeauthor{banerjee2014hierarchical}, \citeyear{banerjee2014hierarchical}, pg.~31) and geometric anisotropy \citep{borgman1994estimation, ecker1999bayesian}. Rapid growth of computing power has allowed the implementation and estimation of these models. 

When modeling a spatial process, the specification of the covariance function will have an effect on kriging and parameter estimates and the associated uncertainty \citep[][pg.~127-135]{cressie1993}. \citet[pg.~87-90]{sherman2011spatial} and \citet{guan2004isotropy} use numerical examples to demonstrate the adverse implications of incorrectly specifying isotropy properties on kriging estimates. Given the variety of choices available regarding the properties of the covariance function (e.g., parametric forms, isotropy, stationarity) and the effect these choices can have on inference, practitioners may seek methods to inform the selection of an appropriate covariance model. 

A number of graphical diagnostics have been proposed to determine isotropy properties. Perhaps the most commonly used methods are directional semivariograms and rose diagrams \citep[pg.~149-154]{matheron1961precision,isaaks1989applied}. \citet[pg.~38-40]{banerjee2014hierarchical} suggest using an empirical semivariogram contour plot to assess isotropy as a more informative method than directional sample semivariograms. Another technique involves comparing empirical estimates of the covariance at different directional lags to assess symmetry for data on gridded sampling locations \citep{modjeska1983spatial}. One caveat of the aforementioned methods is that they can be challenging to assess, are open to subjective interpretations, and can be misleading \citep{guan2004isotropy} because they typically do not include a measure of uncertainty. Experienced statisticians may have intuition about the interpretation and reliability of these diagnostics, but a novice user may wish to evaluate assumptions via a hypothesis test.

\begin{figure}
\begin{center}
\includegraphics[scale = 0.95]{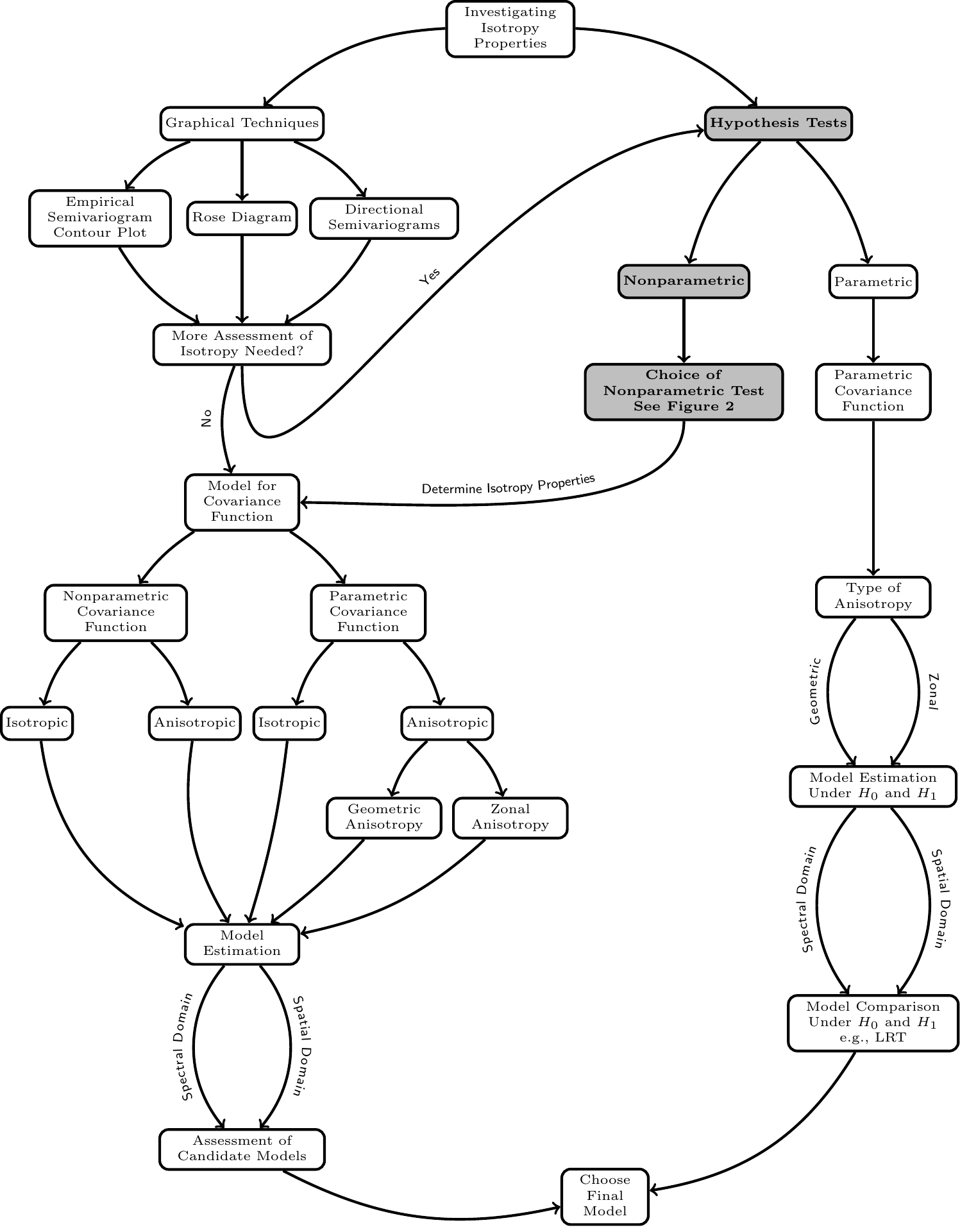}
\end{center}
\caption{A flow chart illustrating the process of determining and modeling isotropy in spatial data. The gray boxes indicate the focus of this paper.}
\label{flowchart1}
\end{figure}

Statistical hypothesis tests of second order properties can be used to supplement and reinforce intuition about graphical diagnostics and can be more objective. Like the graphical techniques, hypothesis tests have their own caveats; for example, a parametric test of isotropy demands specification of the covariance function. A nonparametric method for testing isotropy avoids the potential problems of misspecification of the covariance function and the requirement of model estimation under both the null and alternative hypothesis, which can be computationally expensive for large datasets. Furthermore, nonparametric methods do not require the common assumption of geometric anisotropy. However, in abandoning the parametric assumptions about the covariance function, implementing a test of isotropy presents several challenges (see Section \ref{discussion}). A nonparametric test of isotropy or symmetry can serve as another form of exploratory data analysis that supplements graphical techniques and informs the choice of an appropriate nonparametric or parametric model. Figure \ref{flowchart1} illustrates the process for assessing and modeling isotropy properties.

In this article we review nonparametric hypothesis tests developed to test the assumptions of symmetry and isotropy in spatial processes. We summarize tests in both the spatial and spectral domain and provide tables that enable convenient comparisons of test properties. A simulation study evaluates the empirical size and power of several of the methods and enables a direct comparison of method performance. The simulations also lead to new insights into test performance and implementation beyond those given in the original works. Finally, we include graphics that demonstrate considerations for choosing a nonparametric test and illustrate the process of determining isotropy properties.

The remainder of this article is organized as follows: Section \ref{basicdefs} establishes notation and definitions; Section \ref{isotropy} details the various nonparametric hypothesis tests of isotropy and symmetry and includes Tables \ref{property_table1}-\ref{property_table3} which facilitate comparison between tests as well as test selection for users; Section \ref{simstudy} describes the simulation study comparing the various methods; Sections  \ref{discussion} and \ref{conclusion} provide discussion and conclusions. Additional details on the simulation study are specified in the Appendix.

\section{Notation and Definitions}\label{basicdefs}
Here we briefly review key definitions required for tests of isotropy. For additional background, see \citet{schab}. Let $\{Y(\s): \s \in \D \subseteq \R^d, d > 1\}$ be a second order stationary random field (RF). Below we will assume that $d = 2$, although many of the results hold for the more general case of $d > 2$. For a spatial lag $\h = (h_1, h_2)$, the semivariogram function describes dependence between observatons, $Y$, at spatial locations separated by lag $\h$ and is defined as
\begin{equation}\label{semivariogram}
	\bg(\h) = \frac{1}{2}\mbox{Var}(Y(\s + \h) - Y(\s)).
\end{equation}
  The classical, moment-based estimator of the semivariogram \citep{matheron1962traite} is 
	\begin{equation}\label{classical}
		\widehat{\gamma}(\h) = \frac{1}{2|\D(\h)|} \sum [Y(\s) - Y(\s + \h)]^2,
	\end{equation}
 where the sum is over $\D(\h) = \{\s:\s, \s + \h \in \D\}$ and $|\D(\h)|$ is the number of elements in $\D(\h)$. The set $\D(\h)$ is the set of sampling location pairs that are separated by spatial lag $\h$. The covariance function, $C(\h)$, is an alternative to the semivariogram for describing spatial dependence and is given by $C(\h) = \lim_{||\mathbf{v}|| \rightarrow \infty}\bg(\mathbf{v}) - \bg(\h)$ if the limit exists.
 
Let $\{\s_1, \ldots, \s_n\} \subset \D$ be the finite set of locations at which the random process is observed, providing the random vector $(Y(\s_1), \ldots , Y(\s_n))^T$. The sampling locations may follow one of several spatial sampling designs: gridded locations, randomly and uniformly distributed locations, or a cluster design. Some authors make the distinction between a lattice process and a geostatistical process observed on a grid \citep[pg.~6-10]{fuentes2010spectral, schab}. We do not explore this distinction further and will use the term grid throughout. 

It is often of interest to infer the effect of covariates on the process, deduce dependence structure, and/or predict $Y$ with quantifiable uncertainty at new locations. To achieve these goals, the practitioner must specify the distributional properties of the spatial process. A common assumption is that the finite-dimensional joint distribution is multivariate normal (MVN), in which case we call the RF a Gaussian random field (GRF). We are interested in second order properties; thus, hereafter we assume that the mean of the RF is zero.

A common simplifying assumption on the spatial dependence structure is that it is isotropic.
\begin{definition}\label{iso}
	A second order stationary spatial process is isotropic if the semivariogram, $\bg(\h)$, [or equivalently, the covariance function $C(\h)$] of the spatial process depends on the lag vector $\h$ only through its Euclidean length, $h = ||\h||$, i.e., $\bg(\h) = \bg_{0}(h)$ for some function $\bg_{0}(\cdot)$ of a univariate argument.
\end{definition}
\noindent Isotropy implies that the dependence between any two observations depends only on the distance between their sampling locations and not on their relative orientation. A spatial process that is not isotropic is called anisotropic. Anisotropy is broadly categorized as either geometric or zonal \citep{zimmerman1993another}.  In practice, if a process is assumed to be anisotropic, it is usually assumed to be geometrically anisotropic due to its precise formal and functional definition \citep{ecker1999bayesian}. Geometric anisotropy is governed by a scaling parameter, $R$, and rotation parameter, $\theta$, and implies the anisotropy can be corrected via a linear transformation of the lag vector or, equivalently, the sampling locations \citep[pg.~64]{cressie1993}. 

Symmetry is another directional property of the covariance (semivariogram) function, which is often used to describe the spatial variation of processes on a grid. We discuss symmetry properties here as they are a subset of isotropy, and methods for testing isotropy can often be used to test symmetry. The following definitions come from \citet{lu2005test} and \citet{scaccia2005testing} where the notation $C(h_1, h_2)$ denotes the covariance between random variables located $h_1$ columns and $h_2$ rows apart on a rectangular grid, denoted $\mathcal{L}^2$.

\begin{definition}\label{refsym}
A second order stationary spatial process on a grid is reflection or axially symmetric if $C(h_1, h_2) = C(-h_1, h_2)$ for all $(h_1, h_2) \in \mathcal{L}^2$.
\end{definition}

\begin{definition}\label{diagsym}
A second order stationary spatial process on a grid is diagonally or laterally symmetric if $C(h_1, h_2) = C(h_2, h_1)$ for all $(h_1, h_2) \in \mathcal{L}^2$.
\end{definition}

\begin{definition}\label{comsym}
A second order stationary spatial process on a grid is completely symmetric if it is both reflection and laterally symmetric, i.e., $C(h_1, h_2) = C(-h_1, h_2) = C(h_2, h_1) = C(-h_2, h_1)$ for all $(h_1, h_2) \in \mathcal{L}^2$.
\end{definition}
\noindent Complete symmetry is a weaker property than isotropy. Isotropy requires that $C(h_1,h_2)$ depends only on $\sqrt{h_1^2 + h_2^2}$ for all $h_1, h_2$. The relationship between these properties is given by:

	\begin{equation}\label{relationship}
	\text{isotropy} \implies \text{complete symmetry} \implies
			\begin{array}{l} 
				\text{reflection symmetry} \\ 
				\text{diagonal symmetry}  
			\end{array}.
	\end{equation}
\noindent Thus, rejecting a null hypothesis of reflection symmetry implies evidence against the assumptions of reflection symmetry, complete symmetry, and isotropy. However, failure to reject a null hypothesis of reflection symmetry does not imply an assumption of complete symmetry or isotropy is appropriate.

The aforementioned properties of isotropy and symmetry were defined in terms of examining the spatial random process in the spatial domain, where second order properties depend on the spatial separation, $\h$. Alternatively, a spatial random process can  be represented in the spectral domain using Fourier analysis. For the purposes of investigating second order properties, we are interested in the spectral representation of the covariance function, called the spectral density function and denoted $f(\bo)$, where $\bo = (\omega_1, \omega_2)$. Under certain conditions and assumptions \citep[pg.~62]{fuentes2010spectral}, the spectral density function is given by
\begin{equation}\label{Cinv}
f(\bo) =  \frac{1}{(2\pi)^2}\int_{\R^2} \exp(-i\bo^T\h) C(\h) d\h,
\end{equation}
so that the covariance function, $C(\h)$, and the spectral density function, $f(\bo)$, form a Fourier transform pair. 

Properties of the covariance function imply properties of the spectral density. For example, if the covariance function is isotropic \eqref{iso}, then the spectral density \eqref{Cinv} depends on $\bo$ only through its length, $\omega = ||\bo||$, and we can write $f(\bo) = f_0(\omega)$, where $f_0(\cdot)$ is called the isotropic spectral density \citep{fuentes2013spectral}. Consequently, second order properties of a second order stationary RF can be explored via either the covariance function or the spectral density function. Test statistics quantifying second order properties can be constructed using the periodogram, an estimator of \eqref{Cinv} and denoted by $I(\cdot)$. For a real-valued spatial process on a rectangular grid $\Z^2 \subset \R^2$, a moment-based periodogram used to estimate \eqref{Cinv} is 

\begin{equation}\label{periodogram}
 I(\omega_1, \omega_2) = \frac{1}{(2\pi)^2} \sum_{h_1 = -n_1+1}^{n_1 - 1} \sum_{h_2 = -n_2+1}^{n_2 - 1} \widehat{C}(h_1, h_2) \hspace{1 mm} \cos(h_1\omega_1 + h_2\omega_2), 
\end{equation}
where $n_1$ and $n_2$ denote the number of rows and columns of the grid and $\widehat{C}(h_1, h_2)$ is a non-parametric estimator of the covariance function. In practice, the periodogram \eqref{periodogram} is used to estimate the spectral density at the Fourier or harmonic frequencies. The frequency $\bo = (\omega_1, \omega_2)$ is a Fourier or harmonic frequency if $\omega_j$ is a multiple of $2\pi/n_j$, $j = 1, 2$. Furthermore, the set of frequencies is limited to $\{\omega_j = 2\pi k_j/n_j, k_j = 1, 2, \ldots, n_j^*\}$, where $n_j^*$ is $(n_j - 1)/2$ if $n_j$ is odd and $n_j/2 - 1$ if $n_j$ is even.

\section{Tests of Isotropy and Symmetry}\label{isotropy}
\subsection{Brief History}
\citet{matheron1961precision} developed one of the earliest hypothesis test of isotropy when he used normality of sample variogram estimators to construct a $\chi^2$ test for anisotropy in mineral deposit data. \citet{cabana1987affine} tested for geometric anisotropy using level curves of random fields. \citet{vecchia1988estimation} and \citet{baczkowski1990test} developed tests for isotropy assuming a parametric covariance function. \citet{baczkowski1990testisotropy} also proposed a randomization test for isotropy. Despite these early works, little work on testing isotropy was published during the 1990s, although the PhD dissertation work of \citet{lu1994distributions} would eventually have an noteworthy impact on the literature. Then, in the 2000s, a number of nonparametric tests of second-order properties emerged. Some of the developments used estimates of the variogram or covariogram to test symmetry and isotropy properties \citep{lu2001testing, guan2003nonparametric, guan2004isotropy, guan2007asymptotic, maity2012test}. These works generally borrowed ideas from two bodies of literature: (a) theory on the distributional and asymptotic properties of semivariogram estimators (e.g., \citeauthor{baczkowski1987approximate}, \citeyear{baczkowski1987approximate}; \citeauthor{cressie1993}, \citeyear{cressie1993}, pg.~69-47; \citeauthor{hall1994properties}, \citeyear{hall1994properties}) and (b) subsampling techniques to estimate the variance of statistics derived from spatial data \citep[e.g.,][]{possolo1991subsampling, politis2001moment, sherman1996variance, lahiri2003resampling, lahiri2006resampling}. Other nonparametric methods used the spectral domain to test isotropy and symmetry \citep{ scaccia2002testing,scaccia2005testing,lu2005test, fuentesNStest}. These works generally extended ideas used in the time series literature \citep[e.g.,][]{priestley1969test, priestley1981spectral} to the spatial case. Methods for testing isotropy and symmetry in both the spatial and spectral domains, under the assumption of a parametric covariance function, have also been developed recently \citep{stein2004approximating, haskard2007anisotropic, fuentes2007approximate, matsuda2009fourier, scaccia2011model}.

\subsection{Nonparametric Methods in the Spatial Domain}\label{spatial_domain}
A popular approach to testing second order properties was pioneered in the works of \citet{lu1994distributions} and \citet{lu2001testing} who leveraged the asymptotic joint normality of the sample variogram computed at different spatial lags. The subsequent works of \citet{guan2004isotropy, guan2007asymptotic} and \citet{maity2012test} built upon these ideas and are the primary focus of this subsection. \citet{lu1994distributions} details methods for testing axial symmetry. While \citet{lu2001testing}, \citet{guan2004isotropy}, and \citet{maity2012test} focus on testing isotropy, these methods can also be used to test symmetry. Finally, \citet{bowman2013inference} detail a more computational approach for testing isotropy. Both \citet{li2007nonpar, li2008asymptotic} and \citet{jun2012test} use an approach analogous to the methods from \citet{lu2001testing}, \citet{guan2004isotropy, guan2007asymptotic}, and \citet{maity2012test} but for spatiotemporal data.  Table \ref{property_table1} summarizes test properties discussed in this section and Section \ref{spectral_domain}.

Nonparametric tests for anisotropy in the spatial domain are based on a null hypothesis that is an approximation to isotropy. Under the null hypothesis that the RF is isotropic, it follows that the values of $\bg(\cdot)$ evaluated at any two spatial lags that have the same norm are equal, regardless of the direction of the lags. To fully specify the most general null hypothesis of isotropy, theoretically, one would need to compare variogram values for an infinite set of lags. In practice a small number of lags are specified. Then it is possible to test a hypothesis consisting of a set of linear contrasts of the form 
\begin{equation}\label{linearhypothesis}
H_0: \A\bg(\cdot) = \mathbf{0}
\end{equation}
as a proxy for the null hypothesis of isotropy, where $\A$ is a full row rank matrix \citep{lu2001testing}. For example, a set of lags, denoted $\bL$, commonly used in practice for gridded sampling locations with unit spacing is
\begin{equation}\label{lagset}
 \bL = \{ \h_1 = (1,0), \h_2 = (0,1), \h_3 = (1,1),  \h_4 = (-1,1) \}, 
 \end{equation}
 and the corresponding $\A$ matrix under $H_0: \A\bg(\bL) = \mathbf{0}$ is
\begin{equation}\label{Amat}
	\A = \begin{bmatrix}
		1 & -1 & 0 & 0 \\
		0 & 0 & 1 & -1 
		\end{bmatrix}.
\end{equation}
One of the first steps in detecting potential anisotropy is the choice of lags, as the test results will only hold for the particular set of lags considered \citep{guan2004isotropy}. While this choice is subjective, there are several considerations and useful guidelines for determining the set of lags (see Section \ref{discussion}).

For nonparametric tests of symmetry, the null hypothesis of symmetry using \eqref{linearhypothesis} can be expressed by a countable set of contrasts for a process observed on a grid. Tests of symmetry will be subject to similar practical considers as tests of isotropy, and practitioners testing symmetry properties will need to choose a small set of lags and form a hypothesis that is a surrogate for symmetry. For example, testing reflection symmetry of a process observed on the integer grid would require comparing estimates of $C(\cdot)$ evaluated at the lag pairs $\{(1,0), (-1,0)\}, \{(2,0), (-2,0)\}, \{(1,1), (-1,1)\}$, etc. 

\begin{sidewaystable}

\begin{center}
\caption{Properties of nonparametric tests of isotropy.  ``Domain" refers to the domain used to represent the RF (spatial or spectral),``Test Stat Based On" lists the nonparametric estimator used to construct the test statistic ``Distb'n" gives the limiting asymptotic distribution of the test statistic, and ``GP" denotes whether the test requires data to be generated from a Gaussian process.}
\begin{tabular}{|l||c|c|c|c|c|c|c| } 
\hline 
& \multicolumn{7}{c}{\textbf{Hypothesis Test Properties}}   \\
\hline
 \textbf{Test Method} & Isotropy  & Symmetry & Domain & Test Stat Based On & Asymptotics & Distb'n & GP \\ 
  \hline
 \citet{lu2001testing} & yes  & yes & spatial  & semivariogram  & inc domain &  $\chi^2$ & yes \\ 
 \hline
 \citet{guan2004isotropy, guan2007asymptotic} & yes & yes & spatial   & (kernel)$^a$ variogram  & inc domain & $\chi^2$ $^b$  & no \\ 
  \hline
\citet{scaccia2002testing,scaccia2005testing} & partial & yes  & spectral  & periodogram  & inc domain  &  $Z$, $t$ & no\\ 
 \hline
\citet{lu2005test} & partial & yes  & spectral & periodogram  & inc domain  &  $\chi^2$, $F$ & no\\ 
 \hline
 \citet{fuentesNStest} & partial & no  & spectral & spatial periodogram & shrinking (mixed) &  $\chi^2$ & yes\\
 \hline
 \citet{maity2012test} & yes & yes & spatial  & kernel covariogram& inc domain &  $\chi^2$ & no \\
 \hline
 \citet{bowman2013inference} & yes & no & spatial  & variogram  & inc domain & approx $\chi^2$ & yes \\
 \hline
 \citet{vanhala2014frequency} & yes & yes & spectral  & empirical likelihood  & shrinking (mixed) & $\chi^2$ & no \\
 \hline
  \multicolumn{8}{l}{\footnotesize $^a$for gridded sampling locations, the estimator in \eqref{classical} is used while a kernel variogram estimator is used for non-gridded sampling locations} \\
  \multicolumn{8}{l}{\footnotesize $^b$p-values may need to be approximated using finite sample adjustments} \\
\end{tabular}
\label{property_table1}
\end{center}

\begin{center}
\caption{Test implementation, part 1. ``Subsamp" defines whether spatial subsampling procedures are needed to perform the test, ``S\&P sim" denotes whether or not the author(s) of the method provide a simulation of test size and power (See also Table \ref{property_table3}).}
\begin{tabular}{| l | | c | c | c | c | c | } 
\hline 
& \multicolumn{5}{c}{\textbf{Hypothesis Test Implementation}}   \\
\hline
 \textbf{Test Method} & Sampling Domain Shape & Sampling Design &  Subsamp  & S\&P Sim  & Software \\ 
 \hline
 \citet{lu2001testing} & rectangular in $\mathbb{R}^2$ &  grid  & no  & yes$^a$  &  no  \\ 
 \hline
 \citet{guan2004isotropy, guan2007asymptotic} & convex subsets in $\mathbb{R}^d$ &  grid/unif$^b$/non-unif$^c$  & yes  & yes$^a$ &  R package \texttt{spTest}  \\ 
 \hline
\citet{scaccia2002testing,scaccia2005testing} & rectangular in $\mathbb{R}^2$  & grid  & no  & yes$^a$  & no   \\ 
 \hline
\citet{lu2005test} & rectangular in $\mathbb{R}^2$  & grid  & no   & yes  & R package \texttt{spTest}   \\ 
 \hline
 \citet{fuentesNStest} & rectangular in $\mathbb{R}^d$  & grid  & no  & yes$^a$  & no   \\
 \hline
 \citet{maity2012test} & convex subsets in $\mathbb{R}^d$  &  non-unif$^c$ & yes    & yes$^a$ & R package \texttt{spTest}  \\
 \hline
\citet{bowman2013inference} & convex subsets in $\mathbb{R}^d$  &  unif$^b$  & no    & yes$^a$  & R package \texttt{sm}      \\
 \hline
\citet{vanhala2014frequency} & subsets in $\mathbb{R}^d$  & non-unif$^c$  & no    & yes$^a$  & no      \\
 \hline
  \multicolumn{6}{l}{\footnotesize $^a$ simulated data are Gaussian only} \\
  \multicolumn{6}{l}{\footnotesize $^b$ sampling locations must be generated by homogeneous Poisson process, i.e. uniformly distributed on the domain} \\
  \multicolumn{6}{l}{\footnotesize $^c$ sampling locations can be generated by any general sampling design} \\
\end{tabular}
\label{property_table2}
\end{center}
\end{sidewaystable}
\normalsize
 
 The tests in \citet{lu2001testing}, \citet{guan2004isotropy, guan2007asymptotic}, and \citet{maity2012test} involve estimating either the semivariogram, $\bg(\cdot)$, or covariance, $C(\cdot)$, function at the set of chosen lags, $\bL$. Denoting the set of point estimates of the semivariogram/covariance function at the given lags as $\widehat{\G}_n$, the true values as $\G$, and normalizing constant $a_n$, a central result for all three methods is that
\begin{equation}\label{asymptoticMVN}
a_n(\widehat{\G}_n - \G) \xrightarrow[n \rightarrow \infty]{d} MVN(\mathbf{0}, \mathbf{\Sigma}),
\end{equation}
under increasing domain asymptotics and mild moment and mixing conditions on the RF. Using the $\A$ matrix, an estimate of the variance covariance matrix, $\widehat{\mathbf{\Sigma}}$, and $\widehat{\G}_n$, a quadratic form is constructed, and a p-value can be obtained from an asymptotic $\chi^2$ distribution with degrees of freedom given by the row rank of $\A$. The primary differences between these works are the assumed distribution of sampling locations, the shape of the sampling domain, and the estimation of $\G$ and $\mathbf{\Sigma}$. These differences are important when choosing a test that is appropriate for a particular set of data (see Tables \ref{property_table1} and \ref{property_table2} and Figure \ref{nonpar_choices} for more information about these differences).

\citet{maity2012test} develop a test with the fewest restrictions on the shape of the sampling region and distribution of sampling locations. Their test can be used when the sampling region is any convex subset in $\mathbb{R}^d$ and the distribution of sampling locations in the region follows any general spatial sampling design. The test in \citet{guan2004isotropy} also allows convex subsets in $\mathbb{R}^d$ and is developed for gridded and non-gridded sampling locations but requires non-gridded sampling locations to be uniformly distributed on the domain, i.e., generated by a homogenous Poisson process. The Poisson assumption is dropped in \citet{guan2007asymptotic}. \citet{lu2001testing} require the domain to be rectangular and the observations to lie on a grid.

Another difference between methods is the form of the nonparametric estimator of $\G$. In \citet{lu2001testing}, $\widehat{\G}_n$ is computed using the log of the classical sample semivariogram estimator \eqref{classical}. \citet{guan2004isotropy, guan2007asymptotic} also use the estimator in \eqref{classical} for gridded sampling locations, but use a kernel estimator of $\gamma(\h)$ for non-gridded locations. \citet{maity2012test} use a kernel estimator of the covariance function. When smoothing over spatial lags in $\mathbb{R}^2$, the kernel is typically given as Nadaraya-Watson \citep{nadaraya1964estimating, watson1964smooth} product kernel, independently smoothing over horizontal and vertical lags. Common choices for the kernel are the Epanechnikov or truncated Gaussian kernels. The kernel estimators also require the choice of a bandwidth parameter, $w$. Choosing an appropriate bandwidth is one of the challenges of implementing the tests for non-gridded sampling locations, and the conclusion of the test has the potential to be sensitive to the choice of the bandwidth parameter (see Section \ref{discussion} for recommendations on bandwidth selection).

Nonparametric tests in the spatial domain also vary in the estimation of $\mathbf{\Sigma}$, the asymptotic variance-covariance of $\widehat{\G}_n$ in \eqref{asymptoticMVN}. \citet{lu2001testing} use a plug-in estimator, which requires the choice of a parameter, $m$, that truncates the sum used in estimation. Spatial resampling methods are another approach used to estimate $\Sigma$. The method used for spatial resampling and properties of estimators computed from spatial resampling will depend on the underlying spatial sampling design \citep[pg.~281]{lahiri2003resampling}. \citet{guan2004isotropy, guan2007asymptotic} use a moving window approach, creating overlapping subblocks that cover the sampling region. \citet{maity2012test} employ the grid based block bootstrap (GBBB) \citep{lahiri2006resampling}. The GBBB approach divides the spatial domain into regions, then replaces each region by sampling (with replacement) a block of the sampling domain having the same shape and volume as the region, creating a spatial permutation of blocks of sampling locations. When using the resampling methods, the user must chose the window or block size and the conclusion of the test has the potential to change based on the chosen size. Irregularly shaped sampling domains can pose a challenge in using the subsampling methods. For example, for an irregularly shaped sampling domain, many incomplete blocks may complicate the subsampling procedure. We summarize guidelines for choosing the window/block size in Section \ref{discussion}.

Another approach to testing isotropy in the spatial domain is given by \citet{bowman2013inference} who take a more empirical and computationally-intensive approach. Their methods are available in the \texttt{R} software \citep{rsoftware} package \texttt{sm} \citep{smpackage}. One caveat of using the \texttt{sm} package is that  the methods are computationally expensive, even for moderate sample sizes. For example, a test of isotropy on 300 uniformly distributed sampling locations on a 10 $\times$ 16 sampling domain took approximately 9.5 minutes where the methods from \citet{guan2004isotropy} took 1.6 seconds using a laptop with 8 GB of memory and a 2 GHz Intel Core i7 processor. Because of the computational costs, we do not consider this method further.

\subsection{Nonparametric Methods in the Spectral Domain}\label{spectral_domain}
For gridded sampling locations, nonparametric spectral methods have been developed for testing symmetry \citep{scaccia2002testing, scaccia2005testing,lu2005test} and stationarity \citep{fuentesNStest}, but none are designed with a primary goal of testing isotropy. Due to the difficulties presented by non-gridded sampling locations, historically there have been fewer developments using spectral methods for non-gridded sampling locations than for gridded (or lattice) data, but this is an area that has received more attention recently \citep[see, e.g.,][]{fuentes2007approximate, matsuda2009fourier, band2015frequency}. Despite the challenges, \citet{vanhala2014frequency} have proposed a nonparametric, empirical likelihood approach to test isotropy and separability for non-gridded sampling locations.

The primary motivation for using the spectral domain over the spatial domain are simpler asymptotics in the spectral domain. Unlike estimates of the variogram or covariogram at different spatial lags, estimates of the spectral density at different frequencies via the periodogram are asymptotically independent under certain conditions \citep[pg.~78,194]{pagano1971some, schab}. Additionally, in practice, tests of symmetry in the spectral domain are generally not subject to as many choices (e.g., spatial lag set, bandwidth, block size) as those in the spatial domain. 

Analogous to testing isotropy in the spatial domain by using a finite set of spatial lags, tests of symmetry in the spectral domain typically involve estimating and comparing the spectral density \eqref{Cinv} at a finite set of the Fourier frequencies. For example, axial symmetry \eqref{refsym} of the covariance function implies axial symmetry of the spectral density, $f(\omega_1, \omega_2) = f(-\omega_1, \omega_2)$, which can be evaluated by comparing $I(\omega_1, \omega_2)$ to $I(-\omega_1, \omega_2)$ at a finite set of frequencies. Similarly, the null hypothesis of isotropy can be approximated by comparing periodogram \eqref{periodogram} estimates at a set of distinct frequencies with the same norm \citep{fuentesNStest}. Although most of the current spectral methods are not directly designed to test isotropy, the hypothesis of complete symmetry can be used to reject the assumption of isotropy due to  \eqref{relationship}. However, certain types of anisotropy may not be detected by these tests. For example, a geometrically anisotropic process having the major axis of the ellipses of equicorrelation parallel to the $x$-axis is axially symmetric, and the anisotropy wouldn't be detected by a test of axial symmetry. 

\citet{scaccia2002testing, scaccia2005testing} use the periodogram \eqref{periodogram} to develop a test for axial symmetry. They propose three test statistics that are a function of the periodogram values. The first uses the average of the difference in the log of the periodogram values, $\log[I(\omega_1, \omega_2)] - \log[I(\omega_1, -\omega_2)]$. The second and third test statistics use the average of standardized periodogram differences, $[I(\omega_1, \omega_2) - I(\omega_1, -\omega_2)]/[I(\omega_1, \omega_2) + I(\omega_1, -\omega_2)]$. These test statistics are shown to asymptotically follow a standard normal or $t$ distribution via the Central Limit Theorem, and the corresponding distributions are used to obtain a p-value.

 \citet{lu2005test} also use the periodogram as an estimator of the spectral density to test properties of axial and complete symmetry of processes on the integer grid, $\mathbb{Z}^2$. They use the asymptotic distribution of the periodogram to construct two potential test statistics. Both test statistics leverage the fact that, under certain conditions and at certain frequencies, 
\begin{equation}\label{asymptotics}
	\frac{2I(\omega_1,\omega_2)}{f(\omega_1, \omega_2)} \xrightarrow[n_1,n_2 \rightarrow \infty]{\text{iid}} \chi^2_2.
\end{equation}
Under the null hypothesis of axial or complete symmetry, \eqref{asymptotics} implies that ratios of periodogram values at different frequencies follow an $F(2,2)$ distribution. The preferred test statistic produces a p-value via a Cram\'er-von Mises (CvM) goodness of fit test using the appropriate set of periodogram ratios. Because rejecting a hypothesis of axial symmetry implies rejecting a hypothesis of complete symmetry, \citet{lu2005test} recommend a two-stage procedure for testing complete symmetry. At the first stage, they test the hypothesis of axial symmetry, and if the null hypothesis is not rejected, they test the hypothesis of complete symmetry. To control the overall type-I error rate at $\alpha$, the tests at each stage can be performed using a significance level of $\alpha/2$. 

Leveraging the asymptotic independence of the periodogram at different frequencies, \citet{vanhala2014frequency} propose a spatial frequency domain empirical likelihood (SFDEL) approach that can be used for inference about spatial covariance structure. One of the applications of this method is testing isotropy. An advantage of this method over other frequency domain approaches is that it can be used for non-gridded sampling locations. To implement the test, the user must select the set of lags and, because the sampling locations are not gridded, the number and spacing of frequencies. \citet{vanhala2014frequency} offer some guidelines for these choices based on the simulations and theoretical considerations (e.g., the frequencies need to be asymptotically distant). Once these choices are made, \citet{vanhala2014frequency} maximize an empirical likelihood under a moment constraint corresponding to isotropy and show that the log-ratio of the constrained and unconstrained maximizer asymptotically follows a $\chi^2$ distribution.  The SFDEL method relies on the asymptotic independence of the periodogram values, and the smallest sample size used in simulations was $n = 600$. Thus, it is not clear how the method will perform for smaller sample sizes.
 
\citet{fuentesNStest} introduces a nonparametric, spatially varying spectral density to represent nonstationary spatial processes. While the method can be used to test the assumption of isotropy, the test requires a large sample size on a fine grid. For this reason and also because the test was primarily designed to test the assumption of stationarity, we do not consider it further.

\begin{sidewaystable}
\begin{center}
\caption{Test Implementation, part 2. This table continues the list of choices and considerations for implementing a given test. ``Samp Size (S/A)" indicates the minimum sample sizes used in simulations (S) and applications (A) provided by the author(s) of the method.}
\begin{tabular}{ | l | | >{\raggedright}p{5.5 cm} | >{\raggedright}p{5.5 cm} | c |  } 
\hline 
& \multicolumn{3}{c}{\textbf{Hypothesis Test Implementation}}   \\
\hline
 \textbf{Test Method} & Choices & Other Considerations & Samp Size (S/A)   \\ 
  \hline
 \citet{lu2001testing} & spatial lag set, truncation parameter & optimal truncation parameter   & 100/112        \\ 
 \hline
\citet{guan2004isotropy} gridded design & spatial lag set, window size & optimal window size, edge effects, finite sample adjustment  & 400/289       \\ 
  \hline
   \pbox{20 cm}{ \citet{guan2004isotropy} uniform design \\  \citet{guan2007asymptotic} non-uniform design } & spatial lag set, kernel function, bandwidth parameter, window size & optimal bandwidth \& window size, edge effects, finite sample adjustment  &  \pbox{10 cm}{400/289  \\ 500/584}     \\ 
  \hline
\citet{scaccia2002testing,scaccia2005testing} & test statistic & requires gridded sampling locations; designed to test symmetry   & 121/--     \\ 
 \hline
\citet{lu2005test} & test statistic  & requires gridded sampling locations; two-stage testing procedure, designed to test symmetry; relies on asymptotic independence & 100/--     \\ 
 \hline
 \citet{fuentesNStest} & kernel function, bandwidth parameters, frequency set, spatial knots & requires fine grid; designed to test stationarity   & 5175/5175        \\
 \hline
 \citet{maity2012test} &  lag set $\bL$, kernel function, bandwidth parameter, subblock size, number of bootstrap samples & optimal bandwidth \& block size  & 350/584      \\
 \hline
  \citet{bowman2013inference} & bandwidth parameter   & computationally intensive   & 49/148         \\
 \hline
  \citet{vanhala2014frequency} & lag set, number and spacing of frequencies   & optimal number and spacing of frequencies, relies on asymptotic independence   &  600/--         \\
 \hline
\end{tabular}
\label{property_table3}
\end{center}

\end{sidewaystable}

\normalsize

\section{Simulation Study}\label{simstudy}
We designed a simulation study to compare the empirical size, power, and computational costs for four of the methods. For gridded sampling locations, we compare \citet{lu2005test}[hereafter, \citetalias{lu2005test}] to \citet{guan2004isotropy}[hereafter denoted as \citetalias{guan2004isotropy} or \citetalias{guan2004isotropy}-g when we are specifically referring to the test when applied to gridded sampling locations]. For uniformly distributed sampling locations we compare \citet{maity2012test}[\citetalias{maity2012test}] to \citet{guan2004isotropy, guan2007asymptotic}[\citetalias{guan2004isotropy}-u for the method used for uniformly distributed sampling locations].


We performed the tests on the same realizations of the RF. Data are simulated on rectangular grids or rectangular sampling domains because they are more realistic than square domains and simulations on rectangular domains were not previously demonstrated. We simulate Gaussian data with mean zero and exponential covariance functions with no nugget, a sill of one, and effective range values corresponding to short, medium, and long range dependence. We introduce varying degrees of geometric anisotropy via coordinate transformations governed by a rotation parameter $\theta$ and scaling parameter $R$ that define the ellipses of equicorrelation (see Figure \ref{contours} in the Appendix). The parameter $\theta$ quantifies the angle between the major axis of the ellipse and the $x$-axis (counter-clockwise rotation) while $R$ gives the ratio of the major and minor axes of the ellipse. We also performed simulations that investigate the effect of the lag set, block size, and bandwidth. Although some simulations are given in the original works, our simulations serve to provide a direct comparison of the effects of changing these values and provide further insight into how to choose them. See the Appendix for additional simulation details and results.

\begin{figure}[h]
\begin{center}
\includegraphics[scale = 0.65]{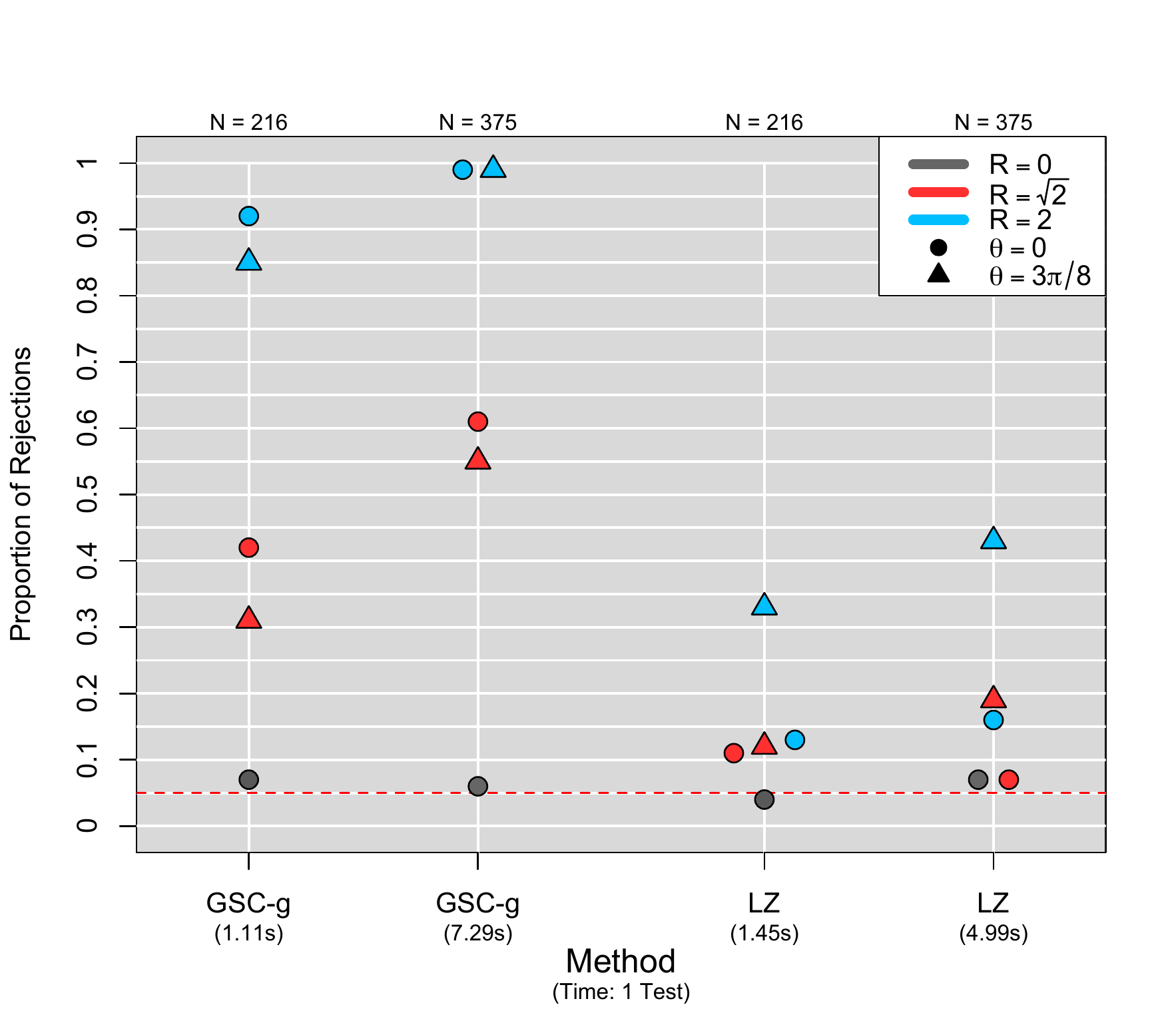}
\end{center}
\caption{Empirical size and power for \citet{guan2004isotropy} [\citetalias{guan2004isotropy}-g] and \citet{lu2005test} [\citetalias{lu2005test}] for 500 realizations of a mean 0 GRF with gridded sampling locations using a nominal level of $\alpha = 0.05$. Colors and shapes indicate the type of anisotropy. Gray points correspond to the isotropic case. The results correspond to a ``medium" effective range. Computational time for each method is also displayed.}
\label{GvL-plot}
\end{figure}

\begin{figure}[h]
\begin{center}
\includegraphics[scale = 0.65]{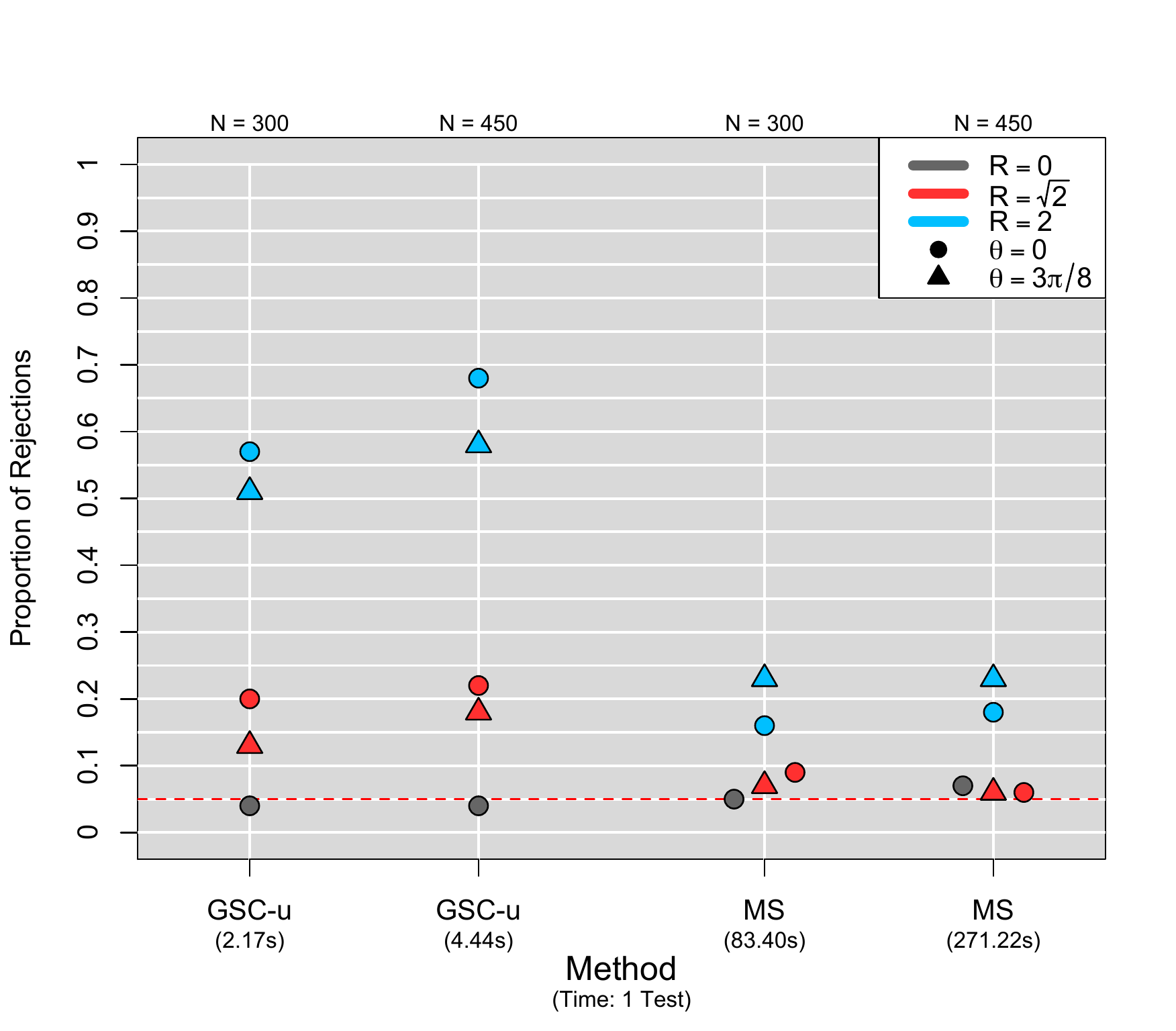}
\end{center}
\caption{Empirical size and power for \citet{guan2004isotropy} [denoted GU] and \citet{maity2012test} [denoted MS] for 200 realizations of a mean 0 GRF with uniformly distributed sampling locations using a nominal level of $\alpha = 0.05$. Colors and shapes indicate the type of anisotropy. Gray points correspond to the isotropic case. The results correspond to a ``medium" effective range. Computational time for each method is also displayed.}
\label{GvM-plot}
\end{figure}

Figures \ref{GvL-plot} and \ref{GvM-plot} illustrate a subset of the simulation results compring empirical size, power, and computational time (full results in Appendix, Tables \ref{GvL} and \ref{GvM}). These simulations indicate that nonparametric tests for anisotropy have higher power for gridded (Figure \ref{GvL}) than for non-gridded (Figure \ref{GvM}) sampling designs. In both comparisons the methods from \citetalias{guan2004isotropy} have favorable empirical power over the competitor with a comparable empirical size. As the effective range increases, both empirical size and power tend to increase for the methods from \citetalias{guan2004isotropy}, but they tend to decrease for \citetalias{maity2012test}. \citetalias{guan2004isotropy}-g and \citetalias{lu2005test} have similar computation time, while \citetalias{maity2012test} is much more computationally expensive than \citetalias{guan2004isotropy}-u. This difference is due to the bootstrapping required by \citetalias{maity2012test}.


Unsurprisingly, as the strength of anisotropy increases (measured by $R$), power increases for all the methods. For a geometrically anisotropic process, the major and minor axes of anisotropy are orthogonal. In comparing the effect of the orientation of isotropy ($\theta$) on the methods, it is important to note that when $\theta = 0$, the major axis of the ellipse defining the geometric anisotropy is parallel to the $x$-axis and corresponds to a spatial process that is axially symmetric but not completely symmetric. When $\theta = 3\pi/8$ the major axis of the ellipse forms a 67.5-degree angle with the $x$-axis, and the spatial process is neither axially nor completely symmetric (see Figure \ref{contours} in the Appendix for contours of equal correlation used in the simulation). The original works generally only simulate data from a geometrically anisotropic process with the major axis of anisotropy forming a 45-degree angle with the $x$-axis; hence, our simulation study more carefully explores the effect of changing the orientation of geometric anisotropy. The methods from \citetalias{guan2004isotropy} exhibit higher power when $\theta = 0$ than when $\theta = 3\pi/8$. This is due to the fact that the lag set, $\bL$, from \eqref{lagset} used for the tests contains a pair of spatial lags that are parallel to the major and minor axes of anisotropy when $\theta = 0$, indicating that an informed choice of spatial lags improves the test's ability to detect anisotropy. This same result does not hold for \citetalias{maity2012test}. It is unclear whether this behavior is observed because the method uses the covariogram rather than the semivariogram, the GBBB rather than moving window approach for estimating $\mathbf{\Sigma}$, or perhaps both. The simulation results indicate that the \citetalias{lu2005test} test has low empirical power; however, this method was developed to test symmetry properties on square grids, and the choice of a rectangular grid for our simulation study does not allow for a large number of periodogram ordinates for the second stage of the procedure for testing the complete symmetry hypothesis. 

Results from simulations that investigate the effects of changing the lag set, the block size, and the bandwidth for non-gridded sampling locations are displayed in Tables \ref{lagset_table}-\ref{bandwidth_table}, respectively, in the Appendix. For both \citetalias{guan2004isotropy}-u and \citetalias{maity2012test}, the lag set in \eqref{lagset} provided an empirical size close to the nominal level. Using more lags or longer lags decreased the size and power for \citetalias{guan2004isotropy}-u. This may be due to the additional uncertainty induced by estimating the covariance between the semivariance at more lags and the larger variance of semivariance estimates at longer lags. For \citetalias{maity2012test} the longer lags lead to an inflated size and more lags decreased the power. In this case, the GBBB may not be capturing the uncertainty in covariance estimates at longer lags with the chosen block size. The \citetalias{maity2012test} test was not overly sensitive to block size with larger blocks leading to slightly higher power. \citetalias{maity2012test} found that an overly large block size was adverse for test size. For \citetalias{guan2004isotropy}-u the small and normal sized windows performed at nominal size levels with comparable power while larger windows were detrimental to test size and power. For \citetalias{guan2004isotropy}-u, we find that choosing a large window tends to lead to overestimation of the asymptotic variance-covariance matrix due to fewer blocks being used to re-estimate the semivariance. Finally, the results investigating the bandwidth selection for \citetalias{guan2004isotropy}-u indicate that choosing an overly large bandwidth inflates test size while choosing too small a bandwidth deflates test size and power. However, the results also indicate that, for the small sample size, test size and power are less negatively affected when approximating the p-value via the finite sample adjustment.

\citet{weller2015sptest} demonstrates applications of several of these methods on two real data sets. The \texttt{R} package \texttt{spTest} \citep{spTest} implements the tests in \citetalias{lu2005test}, \citetalias{guan2004isotropy}, and \citetalias{maity2012test} for rectangular grids and sampling regions and is available on the Comprehensive R Archive Network (\texttt{CRAN}). 


\section{Recommendations}\label{discussion}
\begin{figure}
\begin{center}
\includegraphics[scale = 0.90]{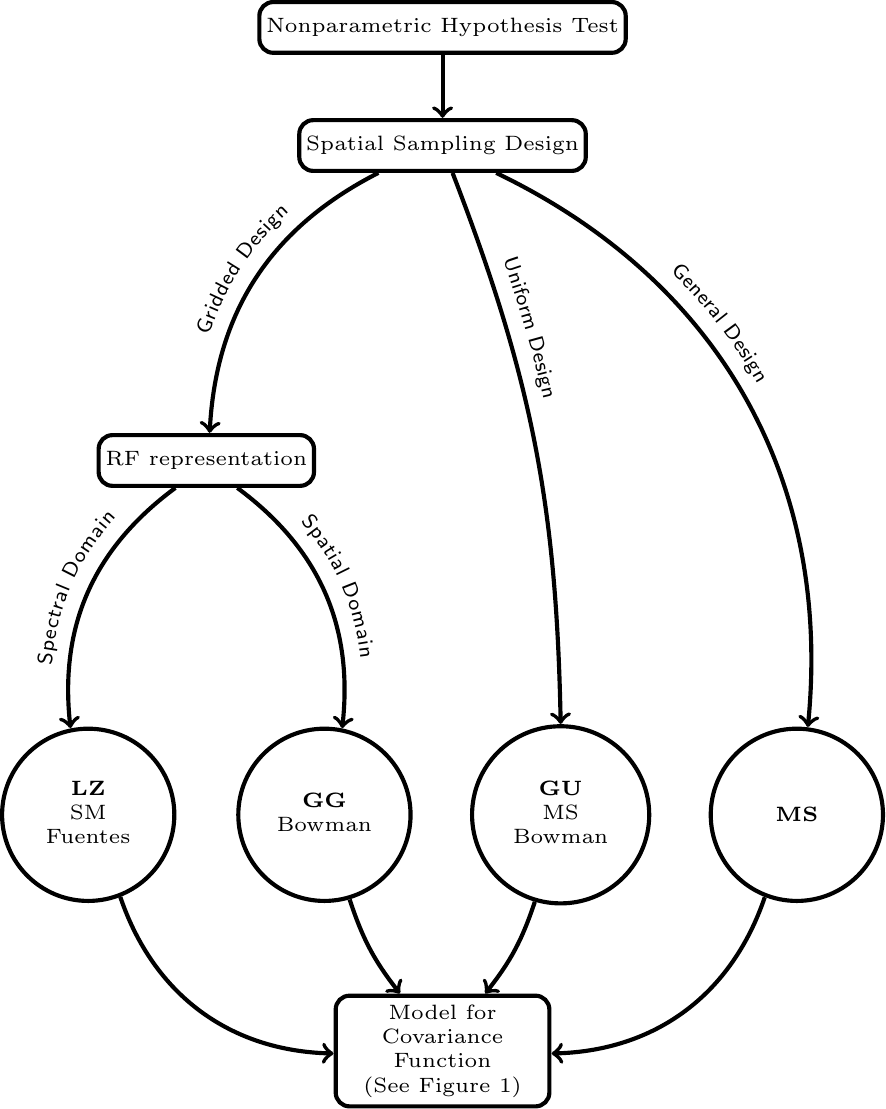}
\end{center}
\caption{Spatial sampling design considerations for choosing a nonparametric hypothesis test of isotropy. The method we recommended for testing isotropy in each situation is given in bold including LZ = \citet{lu2005test}; SM = \citet{scaccia2005testing}; GSC-g = \citet{guan2004isotropy} for gridded sampling locations; GSC-u = \citet{guan2004isotropy} for uniformly distributed sampling locations; MS = \citet{maity2012test}; GSC-n = \citet{guan2007asymptotic} for non-uniform sampling locations.}
\label{nonpar_choices}
\end{figure}

Based on the simulation results we offer recommendations for implementation of nonparametric tests of isotropy. The flow chart in Figure \ref{flowchart1}  along with Figure \ref{nonpar_choices} summarize the steps in the process. Tables \ref{property_table1}-\ref{property_table3} compare the tests. Table \ref{test_impl} summarizes the recommendations provided below.

\begin{sidewaystable}
\begin{center}
\caption{General Recommendations for Test Implementation. This table contains a list of general recommendations for test implementation. These guidelines will not apply in all situations and will vary based on a variety of factors including, but not limited to, the sample size, density of sampling locations, and scale of the problem. See additional discussion in Section \ref{discussion}.}
\begin{tabular}{ | l || l | l | l | >{\raggedright}p{4 cm} | l | } 
\hline 
& \multicolumn{5}{c}{\textbf{Hypothesis Test Choices}}    \\
\hline
 \textbf{Test Method} & Lag Set$^{a}$ & Block Size & Bandwidth & P-value & min.~$n$   \\ 
  \hline
\citet{guan2004isotropy} gridded design & \pbox{25 cm}{ \vspace{0.7 cm} Length: shorter preferred} & $n_b < n^{1/2}$  & n/a  & finite sample adjustment & 150      \\ 

\cline{1-1}\cline{3-6}

   \pbox{20 cm}{ \citet{guan2004isotropy, guan2007asymptotic} uniform design} & \pbox{25 cm}{ \vspace{0.5 cm}Orientation: Eqn \eqref{lagset} }  & $n_b \lesssim n^{1/2}$  & $0.6 < w < 0.9 ^{b}$ & finite sample adjustment when $n < 500$, asymptotic $\chi^2$ when $n \geq 500$ & 300 \\ 

\cline{1-1}\cline{3-6}

 \citet{maity2012test} & \pbox{25 cm}{Number: 4 (2 pairs) \vspace{0.7 cm}} & $n_b \gtrsim n^{1/2}$   & empirical bandwidth  &  asymptotic $\chi^2$  & 300\\
 \hline
 \multicolumn{6}{l}{\footnotesize $^a$ Prior knowledge, if available, should be used to inform the choice of lags.} \\
 \multicolumn{6}{l}{\footnotesize $^b$ Our simulations suggest these bandwidth values are reasonable when using a Gaussian kernel with truncation parameter of 1.5.} \\
  
\end{tabular}
\label{test_impl}
\end{center}
\end{sidewaystable}

In choosing a nonparametric test for isotropy, the distribution of sampling locations on the sampling domain is perhaps the most important consideration. Data on a grid simplifies estimation because the semivariogram or covariogram can be estimated at spatial lags that are exactly observed separating pairs of sampling locations. A grid also allows the option of using easily implemented tests in the spectral domain.

Sample size requirements for the asymptotic properties of tests using the spatial domain to approximately hold will depend on the dependence structure of the random field. \citetalias{guan2004isotropy} note that convergence of their test statistic is slow in the case of gridded sampling locations and obtain an approximate p-value via subsampling rather than the asymptotic $\chi^2$ distribution. Tests using the spectral domain rely on the asymptotic independence of periodogram values, and correlation in finite samples can lead to an inflated test size \citepalias{lu2005test}. Based on our simulations, we recommend the sample size be at least 150 for gridded sampling locations and at least 300 for non-gridded sampling locations. However, power tends be low when the sample size is small and/or the anisotropy is weak (Figures \ref{GvL-plot} and \ref{GvM-plot}).

We focus on implementation of the methods that use the spatial domain for the remainder of this section. We discuss the choice of lags, block size, and bandwidth for the tests in \citetalias{guan2004isotropy} and \citetalias{maity2012test}. Due to the large number of choices required to implement the tests (e.g., block size, bandwidth, kernel function, subsampling method), features of the random field (e.g., sill, range), and properties of the sampling design (e.g., density of sampling locations, shape of sampling domain), the recommendations we offer will not apply in all situations. The numerous moving parts make it challenging to develop general recommendations, especially when choosing a bandwidth. 

When determining the lag set, $\bL$, for use in \eqref{linearhypothesis}, the user needs to select
	\begin{enumerate}[(a)] 
		\item the norm of the lags (e.g., Euclidean distance),
		\item the orientation (direction) of the lags, and
		\item the number of lags. 
	\end{enumerate}	
Regarding (a), short lags are preferred. Estimates of the spatial dependence at large lags may be less reliable than estimates at shorter lags because they are based on a smaller number of pairs of observations and hence more variable. Additionally, empirical and theoretical evidence \citep{lu2001testing} indicates that values of $\bg(\cdot)$ in two different directions generally exhibit the largest difference at a lag less than the effective range, the distance beyond which pairs of observations can be assumed to be independent. Finally, there is mathematical support that correctly specifying the covariance function at short lags is important for spatial prediction \citep{stein1988asymptotically}. Considering (b), if the process is anisotropic, the ideal choice of $\bL$ and $\A$ contrasts lags with the same norm but oriented in the direction of weakest and strongest spatial correlation. Typically, the directions of weakest and strongest spatial correlation will be orthogonal and thus, lags contrasted using the $\A$ matrix should also be orthogonal. Prior information, if available, about the underlying physical/biological process giving rise to the data can also be used to inform the orientation of the lags \citep{guan2004isotropy}. If no prior information about potential anisotropy is available, lags oriented in the same directions as those in \eqref{lagset} are a good starting set. In regards to (c), detecting certain types of anisotropy requires a sufficient number of lags but using a large number of lags requires a large number of observations \citep{guan2004isotropy}. As a general guideline, we suggest using four lags to construct two contrasts.

Several tests require selection of a window or block size to estimate the variance-covariance matrix. The moving window from \citetalias{guan2004isotropy} creates overlapping subblocks of data by sliding the window over a grid placed on the region. Each of these subblocks are used to re-estimate the semivariance. The block size from \citetalias{maity2012test} defines the size of resampled blocks when implementing the GBBB. The GBBB permutes resampled blocks to create a new realization of the process over the entire domain. Choosing the window size in \citetalias{guan2004isotropy} requires balancing two competing goals. First, the moving window should be large enough to create subblocks that are representative of the dependence structure for the entire RF. Second, the window should be small enough to allow for a sufficient number of subblocks to re-estimate the semivariance, as these values are used to obtain an estimate of the asymptotic variance-covariance. A  window that is too large or too small can potentially lead to under- or over-estimation of the asymptotic variance-covariance. For \citetalias{guan2004isotropy}-u, the windows must be large enough to ensure enough pairs of sampling locations are in each subblock to compute an estimate of the semivariance without having to over-smooth. For gridded sampling locations, \citetalias{guan2004isotropy} demonstrate good empirical size and power by using moving windows with size of only $2 \times 2$. However, they find slow convergence to the asymptotic $\chi^2$ distribution, and a p-value is instead computed by approximating the distribution of the test statistic by computing its value for each of the subblocks. Hence, approximating the p-value to two decimal places will require at least 100 subblocks over the sampling region. This may not be possible in practice. For example, a $12 \times 12$ grid of sampling locations with moving windows of size $2 \times 2$ results in only 90 subblocks when correcting for edge effects.  The challenge of choosing the block size in \citetalias{maity2012test} is subject to similar considerations as the window size in \citetalias{guan2004isotropy}. The p-value for both tests will change when performing the test with different window or block sizes, and the user may decide to run the test with different block sizes \citepalias[e.g.,][]{maity2012test}. There are a number of works on resampling spatial data to obtain an estimate of the variance of a spatial statistic \citep[e.g.,][]{sherman1996variance, politis2001moment, lahiri2003resampling, lahiri2006resampling}, but they do not directly consider variance estimation in the case of a nonparametric estimate of the semivariogram/covariogram. Denoting the number of points per block as $n_b$, \citet{sherman1996variance} proposes choosing the block size such that $n_b \approx cn^{1/2}$ for a constant, $c$, when the spatial dependence does not exhibit a large range. In a number of different applications of spatial subsampling, $c$ is typically chosen to be between 0.5 and 2 \citep{politis2001moment, guan2004isotropy, guan2006assessing}. Based on our simulations, we find acceptable empirical size and power for \citetalias{guan2004isotropy}-g using small windows and approximating the p-value with the finite sample adjustment. Thus, we recommend setting $n_b < n^{1/2}$ for \citetalias{guan2004isotropy}-g. For example, we used windows with size $3 \times 2$ and $5 \times 3$ for sampling domains of $18 \times 12$ and $25 \times 15$, respectively. In the case of uniformly distributed sampling locations (see Table \ref{blocksize_table} in the Appendix), the empirical size and power from \citetalias{guan2004isotropy}-u was negatively affected by a large moving window size; hence, we recommend setting $c = 1$ and choosing $n_b \lesssim n^{1/2}$. For the \citetalias{maity2012test} test, a small block size negatively affected the empirical size and power; thus, we recommend choosing $n_b \gtrsim n^{1/2}$ for this test.

Between the choices of a lag set, block size, and bandwidth, choosing an appropriate bandwidth to smooth over observed spatial lags for non-gridded sampling locations is the most challenging. For \citetalias{guan2004isotropy}-u the user needs to choose the form of the smoothing kernel as well as the bandwidth for both the entire grid and the subblocks while \citetalias{maity2012test} use an Epanechnikov kernel and empirical bandwidth based on a user-specified tuning parameter. If the selected bandwidth is too large then over-smoothing occurs. In oversmoothing, there is very little filtering of the lag distance and direction. The lack of filtering produces similar estimates of the spatial dependence at lags with different directions and distances. If the selected bandwidth is too small, then there is very little smoothing and estimates of the spatial dependence are based on a small number of pairs of sampling locations and thus highly variable. Considering the aforementioned effects of the bandwidth, the bandwidth should decrease as $n$ increases under the usual increasing domain asymptotics. For example, simulations (not included) indicated a bandwidth of $w = 0.65$ maintains nominal size when $n = 950$, but leads to deflated test size and power when $n = 400$ on a smaller domain. \citet{garcia2004nonparametric}, \citet{garcia2007asymptotic}, and \citet{kim2012nonparametric} develop theoretically optimal bandwidths for nonparametric semivariogram estimation, but these works are not applicable here because they focus on the isotropic case and require an estimate of the second derivative of the variogram. We have found that the empirical bandwidth used by \citetalias{maity2012test} tends to produce nominal size (see Table \ref{GvM}). For \citetalias{guan2004isotropy}-u we find the most consistent results with a bandwidth in the range of  $0.60 < w < 0.90$ when using a normal kernel truncated at 1.5, but these values will change when a different truncation value or kernel function are employed. For small sample sizes ($n < 500$), our simulations demonstrate that test size and power are less affected by the choice of bandwidth when the p-value is approximated using a finite sample adjustment, indicating poor convergence to the asymptotic $\chi^2$ distribution. Thus, the user should consider using the finite sample adjustment for non-gridded sampling locations when the sample size is small there are at least 100 subblocks.  While it is challenging to choose a bandwidth for \citetalias{guan2004isotropy}-u and the p-value of the test is sensitive to this parameter, the method exhibits nominal size and substantially higher power than \citetalias{maity2012test} when chosen appropriately.

\section{Conclusions}\label{conclusion}
There are several important avenues of future research. Methods to more formally characterize the optimal block size and bandwidth parameters for the tests in the spatial domain would enhance the applicability of the tests. The performance of the tests for non-gridded data in \citet{guan2004isotropy} and \citet{maity2012test} are sensitive to these choices and their optimality remains an open question. \citet{zhang2014self} develop a nonparametric method for estimating the asymptotic variance-covariance matrix of statistics derived from spatial data that avoids choosing tuning parameters which could simplify test implementation. A second area of future work is further development of nonparametric tests of isotropy for gridded and non-gridded data in the spectral domain. A third area of further investigation is to compare nonparametric to parametric methods for testing isotropy, e.g., \citet{scaccia2011model}. A final area of future work is development of a formal definition and more careful quantification of power of the tests. For example, the degree of geometric anisotropy could be quantified using different characteristics of the covariance function, including the ratio of the major and minor axes of the ellipse, degree of rotation of the ellipse from the coordinate axes, and range of the process. Furthermore, it is important to consider the effects of density and design of sampling locations, sample size, and the amount of noise (nugget and sill) in the observations on a test's ability to detect anisotropy.

There is a volume of work on tests for isotropy in other areas of spatial statistics. Methods for detecting anisotropy in spatial point process data have been developed, e.g., \citet[pg.~200-205]{schab}, \citet{guan2003nonparametric}, \citet{guan2006assessing}, and \citet{nicolis2010testing}. For multivariate spatial data, \citet{jona2001modeling} proposes a test for isotropy. \citet{gneiting2007geostatistical} provide a review of potential second order assumptions and models for spatiotemporal geostatistical data, and a number of tests for second order properties of spatiotemporal data have been developed, e.g., \citet{fuentes2006testing}, \citet{li2007nonpar}, \citet{park2008testing}, \citet{shao2009tuning}, \citet{jun2012test}. \citet{li2008testing} construct a test of the covariance structure for multivariate spatiotemporal data. Tests for isotropy have also been developed in the computer science literature \citep[e.g.,][]{molina2002method, chorti2008nonpar, spiliopoulos2011multigrid, thon2015multiscale}.

Appropriately specifying the second order properties of the random field is an important step in modeling spatial data, and a number of models have been developed to capture anisotropy in spatial processes. Graphical tools, such as directional sample semivariograms, are commonly used to evaluate the assumption of isotropy, but these diagnostics can be misleading and open to subjective interpretation. We have presented and reviewed a number of procedures that can be used to more objectively test hypotheses of isotropy and symmetry without assuming a parametric form for the covariance function. These tests may be helpful for a novice user deciding on an appropriate spatial model. In abandoning parametric assumptions, these hypothesis testing procedures are subject and sensitive to choices regarding smoothing parameters, subsampling procedures, and finite sample adjustments. The test that is most appropriate for a set of data will largely depend on the sampling design. Additionally, there are trade-offs between the empirical power demonstrated by the tests and the number of choices user must make to implement the tests (e.g., between \citet{guan2004isotropy} and \citet{maity2012test}). We have offered recommendations regarding the various choices of method and their implementation and have made the tests available in the \texttt{spTest} software. Because of the sensitivity of the tests to the various choices, we believe that graphical techniques and nonparametric hypothesis tests should be used in a complementary role. Graphical techniques can provide an initial indication of isotropy properties and inform sensible choices for a hypothesis test, e.g., in choosing the spatial lag set, while hypothesis tests can affirm intuition about graphical techniques.

\bibliographystyle{apa}
\bibliography{spTestReferences}

\begin{thebibliography}{}

\bibitem[\protect\astroncite{Baczkowski}{1990}]{baczkowski1990testisotropy}
Baczkowski, A. (1990).
\newblock A test of spatial isotropy.
\newblock In {\em Compstat}, pages 277--282. Springer.

\bibitem[\protect\astroncite{Baczkowski and
  Mardia}{1987}]{baczkowski1987approximate}
Baczkowski, A. and Mardia, K. (1987).
\newblock Approximate lognormality of the sample semi-variogram under a
  gaussian process.
\newblock {\em Communications in Statistics-Simulation and Computation},
  16(2):571--585.

\bibitem[\protect\astroncite{Baczkowski and Mardia}{1990}]{baczkowski1990test}
Baczkowski, A. and Mardia, K. (1990).
\newblock A test of spatial symmetry with general application.
\newblock {\em Communications in Statistics-Theory and Methods},
  19(2):555--572.

\bibitem[\protect\astroncite{Bandyopadhyay et~al.}{2015}]{band2015frequency}
Bandyopadhyay, S., Lahiri, S.~N., Nordman, D.~J., et~al. (2015).
\newblock A frequency domain empirical likelihood method for irregularly spaced
  spatial data.
\newblock {\em The Annals of Statistics}, 43(2):519--545.

\bibitem[\protect\astroncite{Banerjee et~al.}{2014}]{banerjee2014hierarchical}
Banerjee, S., Carlin, B.~P., and Gelfand, A.~E. (2014).
\newblock {\em Hierarchical modeling and analysis for spatial data}.
\newblock CRC Press.

\bibitem[\protect\astroncite{Borgman and Chao}{1994}]{borgman1994estimation}
Borgman, L. and Chao, L. (1994).
\newblock Estimation of a multidimensional covariance function in case of
  anisotropy.
\newblock {\em Mathematical geology}, 26(2):161--179.

\bibitem[\protect\astroncite{Bowman and Azzalini}{2014}]{smpackage}
Bowman, A.~W. and Azzalini, A. (2014).
\newblock {\em {R} package \texttt{sm}: nonparametric smoothing methods
  (version 2.2-5.4)}.
\newblock University of Glasgow, UK and Universit\`a di Padova, Italia.

\bibitem[\protect\astroncite{Bowman and Crujeiras}{2013}]{bowman2013inference}
Bowman, A.~W. and Crujeiras, R.~M. (2013).
\newblock Inference for variograms.
\newblock {\em Computational Statistics \& Data Analysis}, 66:19--31.

\bibitem[\protect\astroncite{Cabana}{1987}]{cabana1987affine}
Cabana, E. (1987).
\newblock Affine processes: a test of isotropy based on level sets.
\newblock {\em SIAM Journal on Applied Mathematics}, 47(4):886--891.

\bibitem[\protect\astroncite{Chorti and Hristopulos}{2008}]{chorti2008nonpar}
Chorti, A. and Hristopulos, D.~T. (2008).
\newblock Nonparametric identification of anisotropic (elliptic) correlations
  in spatially distributed data sets.
\newblock {\em Signal Processing, IEEE Transactions on}, 56(10):4738--4751.

\bibitem[\protect\astroncite{Cressie}{1993}]{cressie1993}
Cressie, N. (1993).
\newblock {\em Statistics for Spatial Data: Wiley Series in Probability and
  Statistics}.
\newblock Wiley-Interscience New York.

\bibitem[\protect\astroncite{Ecker and Gelfand}{1999}]{ecker1999bayesian}
Ecker, M.~D. and Gelfand, A.~E. (1999).
\newblock Bayesian modeling and inference for geometrically anisotropic spatial
  data.
\newblock {\em Mathematical Geology}, 31(1):67--83.

\bibitem[\protect\astroncite{Ecker and Gelfand}{2003}]{ecker2003spatial}
Ecker, M.~D. and Gelfand, A.~E. (2003).
\newblock Spatial modeling and prediction under stationary non-geometric range
  anisotropy.
\newblock {\em Environmental and Ecological Statistics}, 10(2):165--178.

\bibitem[\protect\astroncite{Fuentes}{2005}]{fuentesNStest}
Fuentes, M. (2005).
\newblock A formal test for nonstationarity of spatial stochastic processes.
\newblock {\em Journal of Multivariate Analysis}, 96(1):30--54.

\bibitem[\protect\astroncite{Fuentes}{2006}]{fuentes2006testing}
Fuentes, M. (2006).
\newblock Testing for separability of spatial--temporal covariance functions.
\newblock {\em Journal of statistical planning and inference}, 136(2):447--466.

\bibitem[\protect\astroncite{Fuentes}{2007}]{fuentes2007approximate}
Fuentes, M. (2007).
\newblock Approximate likelihood for large irregularly spaced spatial data.
\newblock {\em Journal of the American Statistical Association},
  102(477):321--331.

\bibitem[\protect\astroncite{Fuentes}{2013}]{fuentes2013spectral}
Fuentes, M. (2013).
\newblock Spectral methods.
\newblock {\em Wiley StatsRef: Statistics Reference Online}.

\bibitem[\protect\astroncite{Fuentes and Reich}{2010}]{fuentes2010spectral}
Fuentes, M. and Reich, B. (2010).
\newblock Spectral domain.
\newblock {\em Handbook of Spatial Statistics}, pages 57--77.

\bibitem[\protect\astroncite{Garc{\'\i}a-Soid{\'a}n}{2007}]{garcia2007asymptotic}
Garc{\'\i}a-Soid{\'a}n, P. (2007).
\newblock Asymptotic normality of the {N}adaraya--{W}atson semivariogram
  estimators.
\newblock {\em Test}, 16(3):479--503.

\bibitem[\protect\astroncite{Garc{\'i}a-Soid{\'a}n
  et~al.}{2004}]{garcia2004nonparametric}
Garc{\'i}a-Soid{\'a}n, P.~H., Febrero-Bande, M., and Gonz{\'a}lez-Manteiga, W.
  (2004).
\newblock Nonparametric kernel estimation of an isotropic variogram.
\newblock {\em Journal of Statistical Planning and Inference}, 121(1):65--92.

\bibitem[\protect\astroncite{Gneiting
  et~al.}{2007}]{gneiting2007geostatistical}
Gneiting, T., Genton, M., and Guttorp, P. (2007).
\newblock Geostatistical space-time models, stationarity, separability and full
  symmetry.
\newblock {\em Statistical Methods for Spatio-Temporal Systems}, pages
  151--175.

\bibitem[\protect\astroncite{Guan et~al.}{2004}]{guan2004isotropy}
Guan, Y., Sherman, M., and Calvin, J.~A. (2004).
\newblock A nonparametric test for spatial isotropy using subsampling.
\newblock {\em Journal of the American Statistical Association},
  99(467):810--821.

\bibitem[\protect\astroncite{Guan et~al.}{2006}]{guan2006assessing}
Guan, Y., Sherman, M., and Calvin, J.~A. (2006).
\newblock Assessing isotropy for spatial point processes.
\newblock {\em Biometrics}, 62(1):119--125.

\bibitem[\protect\astroncite{Guan et~al.}{2007}]{guan2007asymptotic}
Guan, Y., Sherman, M., and Calvin, J.~A. (2007).
\newblock On asymptotic properties of the mark variogram estimator of a marked
  point process.
\newblock {\em Journal of statistical planning and inference}, 137(1):148--161.

\bibitem[\protect\astroncite{Guan}{2003}]{guan2003nonparametric}
Guan, Y.~T. (2003).
\newblock {\em Nonparametric methods of assessing spatial isotropy}.
\newblock PhD thesis, Texas A\&M University.

\bibitem[\protect\astroncite{Hall and Patil}{1994}]{hall1994properties}
Hall, P. and Patil, P. (1994).
\newblock Properties of nonparametric estimators of autocovariance for
  stationary random fields.
\newblock {\em Probability Theory and Related Fields}, 99(3):399--424.

\bibitem[\protect\astroncite{Haskard}{2007}]{haskard2007anisotropic}
Haskard, K.~A. (2007).
\newblock {\em An anisotropic Mat\'{e}rn spatial covariance model: REML
  estimation and properties.}
\newblock PhD thesis, University of Adelaide.

\bibitem[\protect\astroncite{Irvine et~al.}{2007}]{irvine2007spatial}
Irvine, K.~M., Gitelman, A.~I., and Hoeting, J.~A. (2007).
\newblock Spatial designs and properties of spatial correlation: effects on
  covariance estimation.
\newblock {\em Journal of agricultural, biological, and environmental
  statistics}, 12(4):450--469.

\bibitem[\protect\astroncite{Isaaks and Srivastava}{1989}]{isaaks1989applied}
Isaaks, E.~H. and Srivastava, R.~M. (1989).
\newblock {\em Applied geostatistics}, volume~2.
\newblock Oxford University Press New York.

\bibitem[\protect\astroncite{Jona-Lasinio}{2001}]{jona2001modeling}
Jona-Lasinio, G. (2001).
\newblock Modeling and exploring multivariate spatial variation: A test
  procedure for isotropy of multivariate spatial data.
\newblock {\em Journal of Multivariate Analysis}, 77(2):295--317.

\bibitem[\protect\astroncite{Journel and Huijbregts}{1978}]{journel1978mining}
Journel, A.~G. and Huijbregts, C.~J. (1978).
\newblock {\em Mining geostatistics}.
\newblock Academic press.

\bibitem[\protect\astroncite{Jun and Genton}{2012}]{jun2012test}
Jun, M. and Genton, M.~G. (2012).
\newblock A test for stationarity of spatio-temporal random fields on planar
  and spherical domains.
\newblock {\em Statistica Sinica}, 22(4):1737.

\bibitem[\protect\astroncite{Kim and Park}{2012}]{kim2012nonparametric}
Kim, T.~Y. and Park, J.-S. (2012).
\newblock On nonparametric variogram estimation.
\newblock {\em Journal of the Korean Statistical Society}, 41(3):399--413.

\bibitem[\protect\astroncite{Lahiri and Zhu}{2006}]{lahiri2006resampling}
Lahiri, S. and Zhu, J. (2006).
\newblock Resampling methods for spatial regression models under a class of
  stochastic designs.
\newblock {\em The Annals of Statistics}, 34(4):1774--1813.

\bibitem[\protect\astroncite{Lahiri}{2003}]{lahiri2003resampling}
Lahiri, S.~N. (2003).
\newblock {\em Resampling methods for dependent data}.
\newblock Springer Science \& Business Media.

\bibitem[\protect\astroncite{Li et~al.}{2007}]{li2007nonpar}
Li, B., Genton, M.~G., and Sherman, M. (2007).
\newblock A nonparametric assessment of properties of space--time covariance
  functions.
\newblock {\em Journal of the American Statistical Association},
  102(478):736--744.

\bibitem[\protect\astroncite{Li et~al.}{2008a}]{li2008testing}
Li, B., Genton, M.~G., and Sherman, M. (2008a).
\newblock Testing the covariance structure of multivariate random fields.
\newblock {\em Biometrika}, 95(4):813--829.

\bibitem[\protect\astroncite{Li et~al.}{2008b}]{li2008asymptotic}
Li, B., Genton, M.~G., Sherman, M., et~al. (2008b).
\newblock On the asymptotic joint distribution of sample space--time covariance
  estimators.
\newblock {\em Bernoulli}, 14(1):228--248.

\bibitem[\protect\astroncite{Lu and Zimmerman}{2001}]{lu2001testing}
Lu, H. and Zimmerman, D. (2001).
\newblock Testing for isotropy and other directional symmetry properties of
  spatial correlation.
\newblock {\em preprint}.

\bibitem[\protect\astroncite{Lu}{1994}]{lu1994distributions}
Lu, H.-C. (1994).
\newblock {\em On the distributions of the sample covariogram and semivariogram
  and their use in testing for isotropy}.
\newblock PhD thesis, University of Iowa.

\bibitem[\protect\astroncite{Lu and Zimmerman}{2005}]{lu2005test}
Lu, N. and Zimmerman, D.~L. (2005).
\newblock Testing for directional symmetry in spatial dependence using the
  periodogram.
\newblock {\em Journal of Statistical Planning and Inference}, 129(1):369--385.

\bibitem[\protect\astroncite{Maity and Sherman}{2012}]{maity2012test}
Maity, A. and Sherman, M. (2012).
\newblock Testing for spatial isotropy under general designs.
\newblock {\em Journal of statistical planning and inference},
  142(5):1081--1091.

\bibitem[\protect\astroncite{Matheron}{1961}]{matheron1961precision}
Matheron, G. (1961).
\newblock Precision of exploring a stratified formation by boreholes with rigid
  spacing-application to a bauxite deposit.
\newblock In {\em International Symposium of Mining Research, University of
  Missouri}, volume~1, pages 407--22.

\bibitem[\protect\astroncite{Matheron}{1962}]{matheron1962traite}
Matheron, G. (1962).
\newblock {\em Trait{\'e} de g{\'e}ostatistique appliqu{\'e}e}.
\newblock Editions Technip.

\bibitem[\protect\astroncite{Matsuda and Yajima}{2009}]{matsuda2009fourier}
Matsuda, Y. and Yajima, Y. (2009).
\newblock Fourier analysis of irregularly spaced data on rd.
\newblock {\em Journal of the Royal Statistical Society: Series B (Statistical
  Methodology)}, 71(1):191--217.

\bibitem[\protect\astroncite{Modjeska and Rawlings}{1983}]{modjeska1983spatial}
Modjeska, J.~S. and Rawlings, J. (1983).
\newblock Spatial correlation analysis of uniformity data.
\newblock {\em Biometrics}, pages 373--384.

\bibitem[\protect\astroncite{Molina and Feito}{2002}]{molina2002method}
Molina, A. and Feito, F.~R. (2002).
\newblock A method for testing anisotropy and quantifying its direction in
  digital images.
\newblock {\em Computers \& Graphics}, 26(5):771--784.

\bibitem[\protect\astroncite{Nadaraya}{1964}]{nadaraya1964estimating}
Nadaraya, E.~A. (1964).
\newblock On estimating regression.
\newblock {\em Theory of Probability \& Its Applications}, 9(1):141--142.

\bibitem[\protect\astroncite{Nicolis et~al.}{2010}]{nicolis2010testing}
Nicolis, O., Mateu, J., and DÕErcole, R. (2010).
\newblock Testing for anisotropy in spatial point processes.
\newblock In {\em Proceedings of the Fifth International Workshop on
  Spatio-Temporal Modelling}, pages 1990--2010.

\bibitem[\protect\astroncite{Pagano}{1971}]{pagano1971some}
Pagano, M. (1971).
\newblock Some asymptotic properties of a two-dimensional periodogram.
\newblock {\em Journal of Applied Probability}, 8(4):841--847.

\bibitem[\protect\astroncite{Park and Fuentes}{2008}]{park2008testing}
Park, M.~S. and Fuentes, M. (2008).
\newblock Testing lack of symmetry in spatial--temporal processes.
\newblock {\em Journal of statistical planning and inference},
  138(10):2847--2866.

\bibitem[\protect\astroncite{Politis and Sherman}{2001}]{politis2001moment}
Politis, D.~N. and Sherman, M. (2001).
\newblock Moment estimation for statistics from marked point processes.
\newblock {\em Journal of the Royal Statistical Society: Series B (Statistical
  Methodology)}, 63(2):261--275.

\bibitem[\protect\astroncite{Possolo}{1991}]{possolo1991subsampling}
Possolo, A. (1991).
\newblock Subsampling a random field.
\newblock {\em Lecture Notes-Monograph Series}, pages 286--294.

\bibitem[\protect\astroncite{Priestley and Rao}{1969}]{priestley1969test}
Priestley, M. and Rao, T.~S. (1969).
\newblock A test for non-stationarity of time-series.
\newblock {\em Journal of the Royal Statistical Society. Series B
  (Methodological)}, pages 140--149.

\bibitem[\protect\astroncite{Priestley}{1981}]{priestley1981spectral}
Priestley, M.~B. (1981).
\newblock {\em Spectral analysis and time series}.
\newblock Academic press.

\bibitem[\protect\astroncite{{R Core Team}}{2014}]{rsoftware}
{R Core Team} (2014).
\newblock {\em R: A Language and Environment for Statistical Computing}.
\newblock R Foundation for Statistical Computing, Vienna, Austria.

\bibitem[\protect\astroncite{Scaccia and Martin}{2002}]{scaccia2002testing}
Scaccia, L. and Martin, R. (2002).
\newblock Testing for simplification in spatial models.
\newblock In {\em Compstat}, pages 581--586. Springer.

\bibitem[\protect\astroncite{Scaccia and Martin}{2005}]{scaccia2005testing}
Scaccia, L. and Martin, R. (2005).
\newblock Testing axial symmetry and separability of lattice processes.
\newblock {\em Journal of Statistical Planning and Inference}, 131(1):19--39.

\bibitem[\protect\astroncite{Scaccia and Martin}{2011}]{scaccia2011model}
Scaccia, L. and Martin, R. (2011).
\newblock Model-based tests for simplification of lattice processes.
\newblock {\em Journal of Statistical Computation and Simulation},
  81(1):89--107.

\bibitem[\protect\astroncite{Schabenberger and Gotway}{2004}]{schab}
Schabenberger, O. and Gotway, C.~A. (2004).
\newblock {\em Statistical methods for spatial data analysis}.
\newblock CRC Press.

\bibitem[\protect\astroncite{Shao and Li}{2009}]{shao2009tuning}
Shao, X. and Li, B. (2009).
\newblock A tuning parameter free test for properties of space--time covariance
  functions.
\newblock {\em Journal of Statistical Planning and Inference},
  139(12):4031--4038.

\bibitem[\protect\astroncite{Sherman}{1996}]{sherman1996variance}
Sherman, M. (1996).
\newblock Variance estimation for statistics computed from spatial lattice
  data.
\newblock {\em Journal of the Royal Statistical Society. Series B
  (Methodological)}, pages 509--523.

\bibitem[\protect\astroncite{Sherman}{2011}]{sherman2011spatial}
Sherman, M. (2011).
\newblock {\em Spatial statistics and spatio-temporal data: covariance
  functions and directional properties}.
\newblock John Wiley \& Sons.

\bibitem[\protect\astroncite{Spiliopoulos
  et~al.}{2011}]{spiliopoulos2011multigrid}
Spiliopoulos, I., Hristopulos, D.~T., Petrakis, M., and Chorti, A. (2011).
\newblock A multigrid method for the estimation of geometric anisotropy in
  environmental data from sensor networks.
\newblock {\em Computers \& Geosciences}, 37(3):320--330.

\bibitem[\protect\astroncite{Stein}{1988}]{stein1988asymptotically}
Stein, M.~L. (1988).
\newblock Asymptotically efficient prediction of a random field with a
  misspecified covariance function.
\newblock {\em The Annals of Statistics}, 16(1):55--63.

\bibitem[\protect\astroncite{Stein et~al.}{2004}]{stein2004approximating}
Stein, M.~L., Chi, Z., and Welty, L.~J. (2004).
\newblock Approximating likelihoods for large spatial data sets.
\newblock {\em Journal of the Royal Statistical Society: Series B (Statistical
  Methodology)}, 66(2):275--296.

\bibitem[\protect\astroncite{Thon et~al.}{2015}]{thon2015multiscale}
Thon, K., Geilhufe, M., and Percival, D.~B. (2015).
\newblock A multiscale wavelet-based test for isotropy of random fields on a
  regular lattice.
\newblock {\em Image Processing, IEEE Transactions on}, 24(2):694--708.

\bibitem[\protect\astroncite{Van~Hala et~al.}{2014}]{vanhala2014frequency}
Van~Hala, M., Bandyopadhyay, S., Lahiri, S.~N., and Nordman, D.~J. (2014).
\newblock A frequency domain empirical likelihood for estimation and testing of
  spatial covariance structure.
\newblock {\em preprint}.

\bibitem[\protect\astroncite{Vecchia}{1988}]{vecchia1988estimation}
Vecchia, A.~V. (1988).
\newblock Estimation and model identification for continuous spatial processes.
\newblock {\em Journal of the Royal Statistical Society. Series B
  (Methodological)}, pages 297--312.

\bibitem[\protect\astroncite{Watson}{1964}]{watson1964smooth}
Watson, G.~S. (1964).
\newblock Smooth regression analysis.
\newblock {\em Sankhy{\=a}: The Indian Journal of Statistics, Series A}, pages
  359--372.

\bibitem[\protect\astroncite{Weller}{2015a}]{spTest}
Weller, Z. (2015a).
\newblock {\em sp{T}est: Nonparametric Hypothesis Tests of Isotropy and
  Symmetry}.
\newblock R package version 0.2.2.

\bibitem[\protect\astroncite{Weller}{2015b}]{weller2015sptest}
Weller, Z.~D. (2015b).
\newblock sp{T}est: an {R} package implementing nonparametric tests of
  isotropy.
\newblock {\em submitted to Journal of Statistical Software, available on
  arXiv}.

\bibitem[\protect\astroncite{Zhang et~al.}{2014}]{zhang2014self}
Zhang, X., Li, B., and Shao, X. (2014).
\newblock Self-normalization for spatial data.
\newblock {\em Scandinavian Journal of Statistics}, 41(2):311--324.

\bibitem[\protect\astroncite{Zimmerman}{1993}]{zimmerman1993another}
Zimmerman, D.~L. (1993).
\newblock Another look at anisotropy in geostatistics.
\newblock {\em Mathematical Geology}, 25(4):453--470.

\end{thebibliography}

\newpage

\setcounter{page}{1}

\section*{Appendix}


\section*{Simulation Study Details and Further Results}\label{simstudy_appendix}
\begin{figure}
\begin{center}
\includegraphics[scale = 0.9]{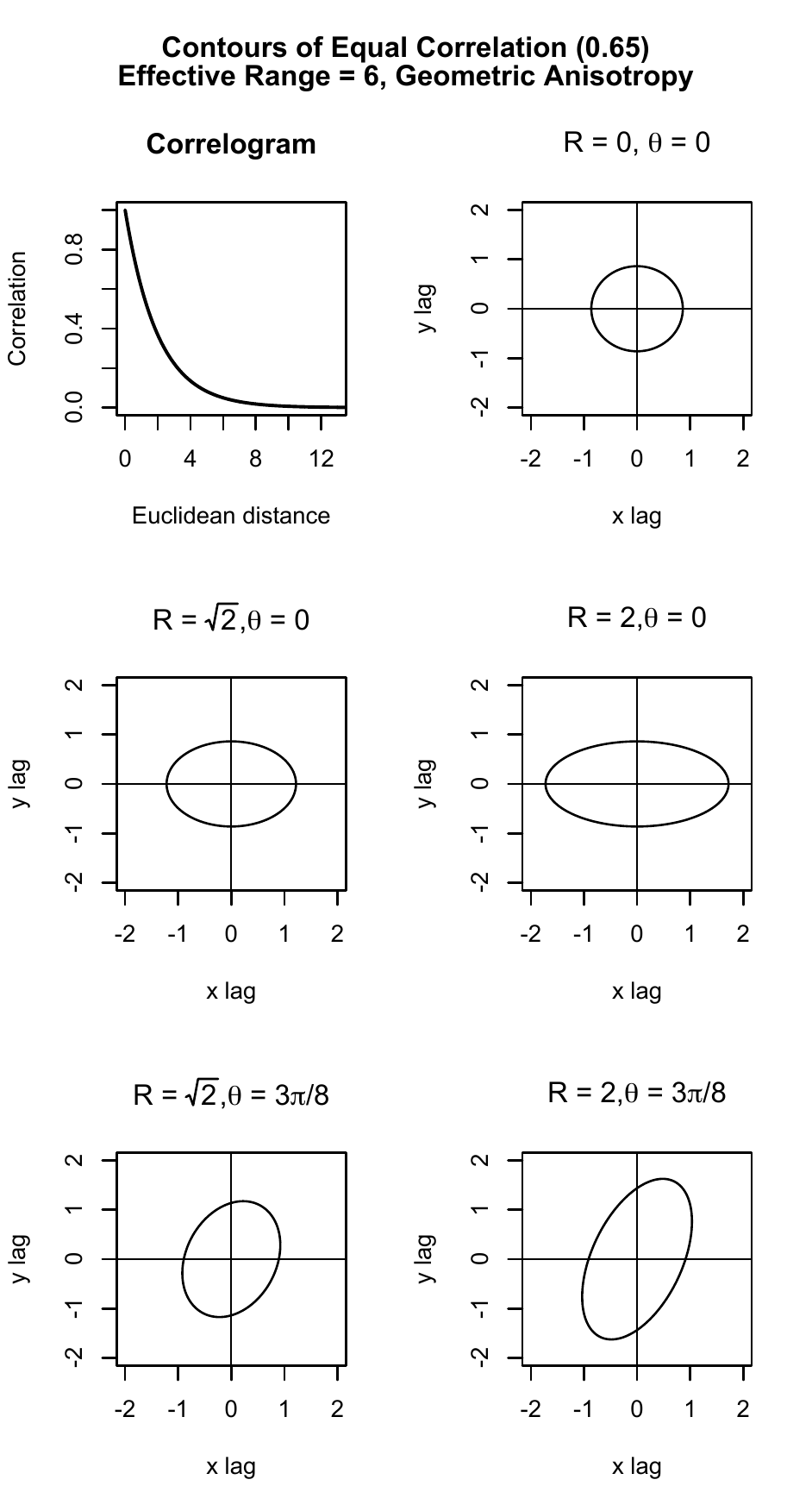}
\end{center}
\caption{Correlogram and contours of equal correlation for the covariance models used in the simulation study.}
\label{contours}
\end{figure}

We define the isotropic exponential covariance function as
	\begin{equation}\label{expcov}
		C(h) =  \left\{ 
			\begin{array}{lr}
				\sigma^2\exp(-\phi h)	  & \text{if } h > 0, \\
				\tau^2 + \sigma^2 & \text{otherwise}
			\end{array}
			\right.
	\end{equation}
 where $h = ||\s_i - \s_j||$ is the distance between sites $\s_i$ and $\s_j$ \citep{irvine2007spatial}. The corresponding semivariogram is $\bg(h) = (\tau^2 + \sigma^2) - \sigma^2\exp(-\phi h)$, where $\tau^2$ is the nugget, $\tau^2 + \sigma^2$ is the sill, and the effective range, $\xi$, the distance beyond which the correlation between observations is less than 0.05, is 
 \[
 	\xi = \frac{-1}{\phi} \log\left(0.05 \frac{\tau^2 + \sigma^2}{\sigma^2}\right).
 \]
Simulations in Section \ref{simstudy} were performed using the exponential covariance function \eqref{expcov} with a partial sill, $\sigma^2$, of 1 and no nugget, $\tau^2 = 0$. We also performed simulations using different nugget values (results not included). As expected, introducing a nugget had an adverse effect on empirical test size and power. For the no nugget simulations, effective ranges, $\xi$, for isotropic processes were chosen to be 3, 6, and 12 corresponding to short, medium, and long range dependence. Geometric anisotropy was introduced by transforming the sampling locations according to a scaling parameter, $R$, and a rotation parameter, $\theta$. Given an $(R, \theta)$ pair, the coordinates $(x,y)$ are transformed to the ``anisotropic" coordinates, $(x_a, y_a)$ via
\[
(x_a, y_a) = (x, y) \begin{bmatrix} \cos\theta & \sin\theta \\ -\sin\theta & \cos\theta \end{bmatrix} \begin{bmatrix} 1 & 0 \\ 0 & \frac{1}{R} \end{bmatrix}.
\]
A realization from the anisotropic process is then created by simulating using the distance matrix from the transformed coordinates and placing the observed values at their corresponding untransformed sampling locations. Figure \ref{contours} shows the isotropic exponential correlogram corresponding to $\tau^2 = 1$ and $\xi = 6$ and contours of equicorrelation corresponding to the $(R, \theta)$ values used in the simulation study. Note that a larger value of $R$ corresponds to a more anisotropic process.

For the simulations comparing the GSC-g and LZ \citep{lu2005test} tests in Table \ref{GvL}, data were simulated on a subset of the integer grid, $\mathbb{Z}^2$. The p-values for the GG test were approximated using a finite sample statistic \citep{guan2004isotropy}, and we used the lag set in \eqref{lagset} and $\A$ matrix in \eqref{Amat}. For the results involving the LZ test, a test of complete symmetry was performed as an approximation to the null hypothesis of isotropy. The p-values for the LZ test were obtained using the CvM* statistic. A nominal level of $\alpha = 0.05$ was maintained by first testing reflection symmetry at $\alpha = 0.025$ then testing complete symmetry at $\alpha = 0.025$ if the hypothesis of reflection symmetry was not rejected. For the GG test, the moving window dimensions were $3 \times 2$ (width, height) and $5 \times 3$ for the parent grids of $18 \times 12$ and $25 \times 15$, respectively.

For the simulations in Table \ref{GvM} comparing the GU \citep{guan2004isotropy} and MS \citep{maity2012test} tests, data were simulated at random, uniformly distributed sampling locations on $10 \times 16$ and $10 \times 20$ sampling domains. The lag set, $\bL$, used for both tests is given in \eqref{lagset} with $\A$ matrix \eqref{Amat}, and the p-values for both methods were obtained using the asymptotic $\chi_2^2$ distribution. For semivariogram estimates in GU, we use independent (product) Gaussian (normal) kernels with a truncation parameter of 1.5. The bandwidth for the Gaussian kernel for smoothing over lags on the entire field and on moving windows was chosen as $w = 0.75$. We used the empirical bandwidth and the product Epanechnikov kernel given in \citet{maity2012test} to implement the MS test. For both tests, a grid with spacing of 1 was laid on the sampling region. Using this grid, the moving window dimensions for the GU test were 4 $\times$ 2 and the block size for the MS test were 4 $\times$ 2. For the MS test, $B = 100$ resamples using the GBBB were used to estimate the asymptotic variance-covariance matrix.

For the results in Tables \ref{lagset_table} - \ref{bandwidth_table}, we simulated mean 0, Gaussian RFs with exponential covariance function with no nugget, a sill of one, and medium effective range ($\xi = 6$). Sampling locations were generated randomly and uniformly over a $16 \times 10$ sampling domain. We use the lag set and $\A$ matrix from \ref{lagset} and \ref{Amat}, respectively, unless otherwise noted. All tests were performed using a nominal level of $\alpha = 0.05$. For the GU tests, we use product Gaussian kernels with a truncation parameter of 1.5. For the MS tests, we use the default product Epanechnikov kernels with empirical bandwidth specified in \citet{maity2012test}.

\begin{figure}
\begin{center}
\includegraphics[scale = 0.55]{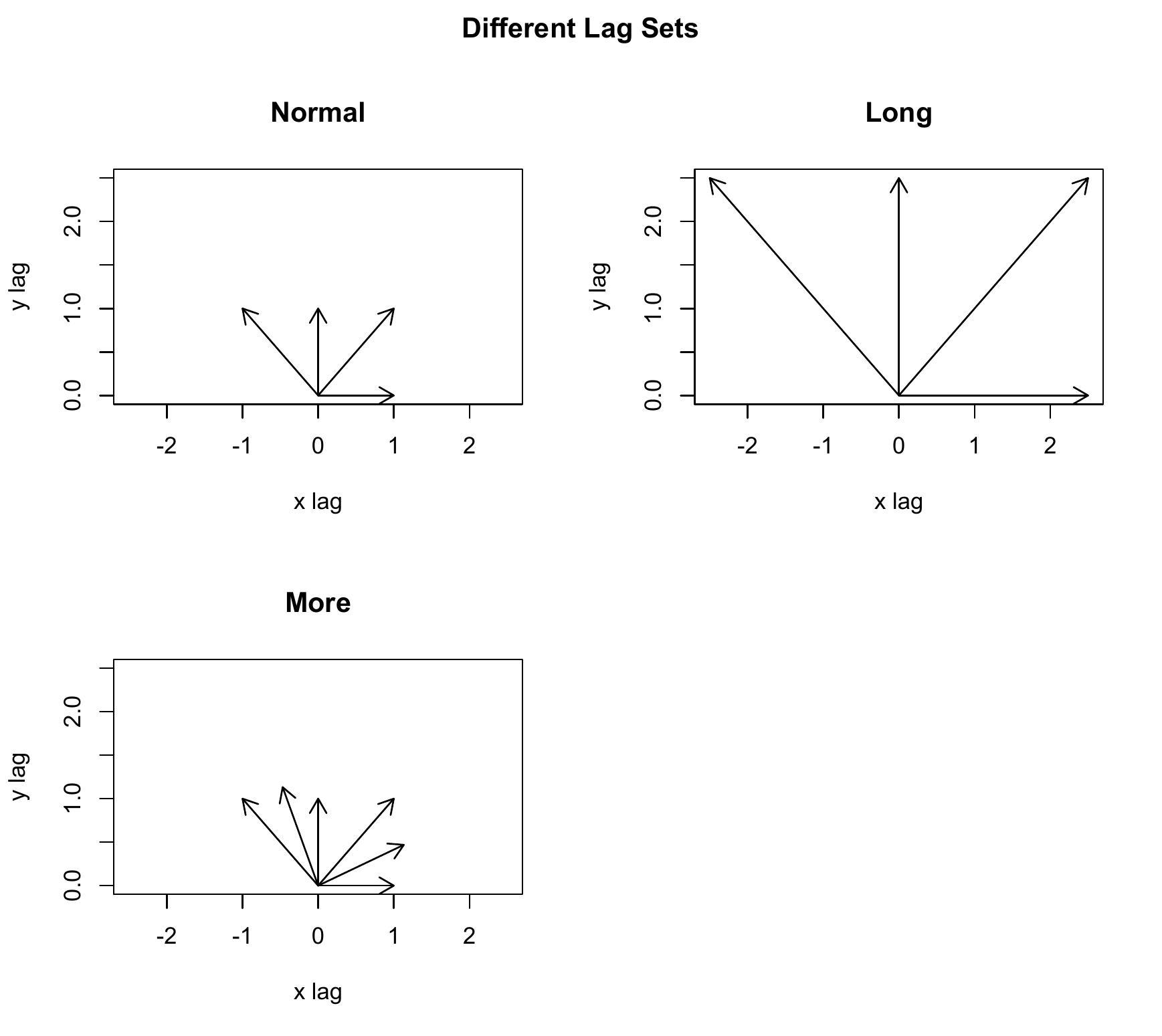}
\end{center}
\caption{The lag sets used for the simulations in Table \ref{lagset_table}.}
\label{lagplots}
\end{figure}

The simulation results in Table \ref{lagset_table} demonstrate the effects of changing the set of lags for the GU and MS tests. For these simulations, the lag set labeled ``normal" corresponds to the lag set given in \eqref{lagset}. The lag set labeled ``long" represents the lags in \eqref{lagset} multiplied by 2.5. Finally, the lag set labeled ``more" stands for the lags in \eqref{lagset} with the additional pair of lags $\{\h_5 = (1.132, 0.469), \h_6 = (-0.469, 1.132)\}$. The lags $\h_5$ and $\h_6$ are a pair of lags the create approximate $22.5^{\circ}$ and $112.5^{\circ}$ angles, respectively, with the $x$-axis (counter-clock wise rotation) and have Euclidean length of approximately 1.22. These were chosen to supplement the lag pairs $(\h_1, \h_2)$ which have unit length and create $0^{\circ}$ and $90^{\circ}$ angles with the $x$-axis and $(\h_3, \h_4)$ which have length $\sqrt{2} \approx 1.41$ and create $45^{\circ}$ and $135^{\circ}$ angles with the $x$-axis. The lag sets are plotted in Figure \ref{lagplots}. The $\A$ matrix for the ``more" lagset was constructed as in \eqref{Amat}, where orthogonal lags are contrasted. The p-values were calculated using the asymptotic $\chi^2$ distribution with degrees of freedom based on the number of pairs of lags contrasted. For the GU method, we used a bandwidth of 0.75. The moving window dimensions were 4 $\times$ 2. For the MS method, we chose block dimensions of 4 $\times$ 2 and used  $B = 75$ resamples using the GBBB to estimate the asymptotic variance-covariance matrix. Table \ref{blocksize_table} demonstrates the effects of changing the block size for the GU and MS tests. For these simulations, the labels ``small", ``normal", and ``large" correspond to moving windows/blocks of size $3 \times 2$, $4 \times 2$, and $5 \times 3$, respectively. Because we simulated $n = 300$ uniformly distributed sampling locations on a $16 \times 10$ domain, we expect 1.875 sampling locations per unit area. Thus, we expect $n_b = $ 11.3, 15, and 28.1 points per block for the small, normal, and large block sizes, respectively. We find that the methods tend to have nominal size when $n_b \approx n^{1/2} = 17.3$. For both tests, we used the lags in \eqref{lagset}, and the blocks are defined by a grid with spacing 0.5 placed on the sampling region (i.e., a $4 \times 2$ window is achieved by setting the window dimensions to $8 \times 4$ in the \texttt{spTest} software). We performed the tests using a nominal level of $\alpha = 0.05$, and the p-values were calculated using the asymptotic $\chi^2$ distribution. For the GU method, we used a bandwidth of 0.75. For the MS method, we used  $B = 100$ resamples using the GBBB to estimate the asymptotic variance-covariance matrix. Finally, Table \ref{bandwidth_table} demonstrates the effects of changing the bandwidth for the GU test. We use bandwidths of $w = $ 0.65, 0.75, and 0.85. The p-values are calculated using both the asymptotic $\chi^2$ distribution and using a finite sample adjustment similar to the one used by \citet{guan2004isotropy} for gridded sampling locations.

\begin{table}[p]
\caption{Empirical size and power for \citet{guan2004isotropy} [denoted GG] and \citet{lu2005test} [denoted LZ] for 500 realizations of a mean 0 GRF with gridded sampling locations using a nominal level of $\alpha = 0.05$. Computational time for each method is also included.}
\begin{subtable}{0.5\linewidth}
\centering
\subcaption{Sample size of $n = 216$ gridded sampling locations.}
\begin{tabular}{|l | l | c |p{6 mm} | p{6 mm}| p{6 mm} |}
\multicolumn{6}{c}{\textbf{18 cols $\times$ 12 rows grid}} \\
\hline
\multicolumn{3}{|c|}{} & \multicolumn{3}{c|}{\textbf{effective range}}  \\
\hline
$R$ & $\theta$ & Method &  3 & 6 & 12 \\
\hline
 \parbox[t][0.75mm][c]{4.0 mm}{0} & \parbox[t][0.75mm][c]{4.0 mm}{0} & GG & 0.05  & 0.07  & 0.05  \\
   &  &     LZ & 0.04  & 0.04  & 0.08   \\
   \hline
 \parbox[t][0.75mm][c]{4.0 mm}{$\sqrt{2}$} & \parbox[t][0.75mm][c]{4.0 mm}{$0$}  & GG & 0.32  & 0.42  & 0.43  \\
   &  &                 LZ & 0.06  & 0.11  & 0.12  \\
   \hline
  \parbox[t][0.75mm][c]{4.0 mm}{2} & \parbox[t][0.75mm][c]{4.0 mm}{0}  & GG & 0.91  & 0.92  & 0.94  \\
   &  &      LZ & 0.14 & 0.13  & 0.15  \\
   \hline
 \parbox[t][0.75mm][c]{4.0 mm}{$\sqrt{2}$} & \parbox[t][0.75mm][c]{4.0 mm}{$\frac{3\pi}{8}$} & GG & 0.27  & 0.31  & 0.34  \\
   &  &                                      LZ & 0.14 & 0.12  & 0.13  \\
   \hline
   \parbox[t][0.75mm][c]{4.0 mm}{2} & \parbox[t][0.75mm][c]{4.0 mm}{$\frac{3\pi}{8}$} & GG & 0.77  & 0.85  & 0.86  \\
   &  &                          LZ & 0.29  & 0.33  & 0.33  \\
   \hline
   \hline
   \multicolumn{6}{|c|}{\textbf{Computational Time for 1 Test}}   \\
   \hline
   \multicolumn{4}{|c|}{GG}  &  \multicolumn{2}{r|}{1.11 seconds}  \\
    \multicolumn{4}{|c|}{LZ}  &   \multicolumn{2}{r|}{1.45 seconds}  \\
   \hline
\end{tabular}
\label{GvL1}
\end{subtable}
\quad
\begin{subtable}{0.5\linewidth}
\centering
\subcaption{Sample size of $n = 375$ gridded sampling locations.}
\begin{tabular}{|l | l | c |p{6 mm} | p{6 mm}| p{6 mm} |}
\multicolumn{6}{c}{\textbf{25 cols $\times$ 15 rows grid}} \\
\hline
\multicolumn{3}{|c|}{} & \multicolumn{3}{c|}{\textbf{effective range}}  \\
\hline
$R$ & $\theta$ & Method &  3 & 6 & 12 \\
\hline
 \parbox[t][0.75mm][c]{4.0 mm}{0} & \parbox[t][0.75mm][c]{4.0 mm}{0} & GG & 0.05 & 0.06  & 0.07  \\
   &  &      LZ & 0.06  & 0.07 & 0.07   \\  
   \hline
 \parbox[t][0.75mm][c]{4.0 mm}{$\sqrt{2}$} & \parbox[t][0.75mm][c]{4.0 mm}{$0$}  & GG & 0.63  & 0.61  & 0.69  \\
   &  &                 LZ & 0.07 & 0.09 & 0.10  \\
   \hline
  \parbox[t][0.75mm][c]{4.0 mm}{2} & \parbox[t][0.75mm][c]{4.0 mm}{0}  & GG & 0.98  & 0.99  & 0.99  \\
   &  &       LZ & 0.14 & 0.16  & 0.15  \\
   \hline
 \parbox[t][0.75mm][c]{4.0 mm}{$\sqrt{2}$} & \parbox[t][0.75mm][c]{4.0 mm}{$\frac{3\pi}{8}$} & GG & 0.47  & 0.55  & 0.55  \\
   &  &                                       LZ & 0.16 & 0.19  & 0.18  \\
   \hline
   \parbox[t][0.75mm][c]{4.0 mm}{2} & \parbox[t][0.75mm][c]{4.0 mm}{$\frac{3\pi}{8}$} & GG & 0.97  & 0.99  & 0.98  \\
   &  &                           LZ & 0.37  & 0.43  & 0.45  \\
   \hline
   \hline
   \multicolumn{6}{|c|}{\textbf{Computational Time for 1 Test}}   \\
   \hline
   \multicolumn{4}{|c|}{GG}  &  \multicolumn{2}{r|}{7.29 seconds}  \\
    \multicolumn{4}{|c|}{LZ}  &   \multicolumn{2}{r|}{4.99 seconds}  \\
   \hline
\end{tabular}
\label{GvL2}
\end{subtable}
\label{GvL}
\end{table}

\begin{table}
\caption{Empirical size and power for \citet{guan2004isotropy} [denoted GU] and \citet{maity2012test} [denoted MS] for 200 realizations of a mean 0 GRF with uniformly distributed sampling locations using a nominal level of $\alpha = 0.05$. Computational time for each method is also included.}
\begin{subtable}{0.5\linewidth}
\centering
\subcaption{Sample size of $n = 300$ uniformly distributed sampling locations.}
\begin{tabular}{| l | l | c |p{6 mm} | p{6 mm}| p{6 mm} |}
\multicolumn{6}{c}{\textbf{10 height $\times$ 16 width domain}} \\
\hline
\multicolumn{3}{|c|}{}  & \multicolumn{3}{c|}{\textbf{effective range}}  \\
\hline
$R$ & $\theta$ & Method &  3 & 6 & 12 \\
\hline
  \parbox[t][0.75mm][c]{4.0 mm}{0} &  \parbox[t][0.75mm][c]{4.0 mm}{0} & GU & 0.02  & 0.04 & 0.05  \\
   &  &   															 MS & 0.04  & 0.05  & 0.04  \\
   \hline
 \parbox[t][0.75mm][c]{4.0 mm}{$\sqrt{2}$} &  \parbox[t][0.75mm][c]{4.0 mm}{0} & GU & 0.15  & 0.20 & 0.27  \\
                  &  &														 MS & 0.10  & 0.09 & 0.08  \\
   \hline
   \parbox[t][0.75mm][c]{4.0 mm}{2} &  \parbox[t][0.75mm][c]{4.0 mm}{0} & GU & 0.43  & 0.57  & 0.62  \\
   &  &    															 MS & 0.21 & 0.16  & 0.15  \\
   \hline
 \parbox[t][0.75mm][c]{4.0 mm}{$\sqrt{2}$} & \parbox[t][0.75mm][c]{4.0 mm}{$\frac{3\pi}{8}$} & GU & 0.12  & 0.13  & 0.16  \\
   &  &                                    															  MS & 0.08 & 0.07  & 0.04  \\
   \hline
  \parbox[t][0.75mm][c]{4.0 mm}{2} & \parbox[t][0.75mm][c]{4.0 mm}{$\frac{3\pi}{8}$} & GU & 0.37  & 0.51  & 0.51  \\
   &  &                        														 MS & 0.27  & 0.23  & 0.21  \\
   \hline
   \hline
   \multicolumn{6}{|c|}{\textbf{Computational Time for 1 Test}}   \\
   \hline
   \multicolumn{4}{|c|}{GU}  &  \multicolumn{2}{r|}{2.17 seconds}  \\
    \multicolumn{4}{|c|}{MS}  &   \multicolumn{2}{r|}{83.40 seconds}  \\
   \hline
\end{tabular}
\label{GvM1}
\end{subtable}
\quad
\begin{subtable}{0.5\linewidth}
\centering
\subcaption{Sample size of $n = 450$ uniformly distributed sampling locations.}
\begin{tabular}{| l | l | c |p{6 mm} | p{6 mm}| p{6 mm} |}
\multicolumn{6}{c}{\textbf{10 height $\times$ 20 width domain}} \\
\hline
\multicolumn{3}{|c|}{}  & \multicolumn{3}{c|}{\textbf{effective range}}  \\
\hline
$R$ & $\theta$ & Method &  3 & 6 & 12 \\
\hline
  \parbox[t][0.75mm][c]{4.0 mm}{0} &  \parbox[t][0.75mm][c]{4.0 mm}{0} & GU & 0.00  & 0.04  & 0.05  \\
   &  &    															MS & 0.05  & 0.07 & 0.03  \\
   \hline
 \parbox[t][0.75mm][c]{4.0 mm}{$\sqrt{2}$} &  \parbox[t][0.75mm][c]{4.0 mm}{0} & GU & 0.15  & 0.22  & 0.23  \\
                  &  & 															MS & 0.07  & 0.06 & 0.07  \\
   \hline
   \parbox[t][0.75mm][c]{4.0 mm}{2} &  \parbox[t][0.75mm][c]{4.0 mm}{0} & GU & 0.57  & 0.68  & 0.75 \\
   &  &     														  MS & 0.32 & 0.18  & 0.14  \\
   \hline
 \parbox[t][0.75mm][c]{4.0 mm}{$\sqrt{2}$} & \parbox[t][0.75mm][c]{4.0 mm}{$\frac{3\pi}{8}$} & GU & 0.09  & 0.18 & 0.21  \\
   &  &                                      															MS & 0.12 & 0.06  & 0.08  \\
   \hline
  \parbox[t][0.75mm][c]{4.0 mm}{2} & \parbox[t][0.75mm][c]{4.0 mm}{$\frac{3\pi}{8}$} & GU & 0.55  & 0.58  & 0.65  \\
   &  &                        												MS & 0.37  & 0.23  & 0.21  \\
   \hline
   \hline
   \multicolumn{6}{|c|}{\textbf{Computational Time for 1 Test}}   \\
   \hline
   \multicolumn{4}{|c|}{GU}  &  \multicolumn{2}{r|}{4.44 seconds}  \\
    \multicolumn{4}{|c|}{MS}  &   \multicolumn{2}{r|}{162.35 seconds}  \\
   \hline
\end{tabular}
\label{GvM2}
\end{subtable}
\label{GvM}
\end{table}

\begin{table}
\caption{Effects of changing the lag set. Empirical size and power for \citet{guan2004isotropy} [denoted GU] and \citet{maity2012test} [\citetalias{maity2012test}] for 100 realizations of a mean 0 GRF with $n = 400$ uniformly distributed sampling locations. The label ``normal" corresponds to the lag set in \eqref{lagset}, while ``long" represents using longer lags, and ``more" denotes using more lags (see Figure \ref{lagplots}).}
\centering
\begin{tabular}{| l | l | c |p{9 mm} | p{9 mm}| p{9 mm} |}
\multicolumn{6}{c}{\textbf{16 width $\times$ 10 height domain}} \\
\hline
\multicolumn{3}{|c|}{}  & \multicolumn{3}{c|}{\textbf{Lag Set}}  \\
\hline
$R$ & $\theta$ & Method &  normal & long & more \\
\hline
  \parbox[t][0.75mm][c]{4.0 mm}{0} &  \parbox[t][0.75mm][c]{4.0 mm}{0} & GU & 0.02  & 0.00 & 0.01  \\
   &  &   															 MS & 0.03  & 0.14  & 0.03  \\
   \hline
 \parbox[t][0.75mm][c]{4.0 mm}{$\sqrt{2}$} & \parbox[t][0.75mm][c]{4.0 mm}{$\frac{3\pi}{8}$} & GU & 0.19  & 0.07  & 0.16  \\
   &  &                                    															  MS & 0.11 & 0.24  & 0.07  \\
   \hline
  \parbox[t][0.75mm][c]{4.0 mm}{2} & \parbox[t][0.75mm][c]{4.0 mm}{$\frac{3\pi}{8}$} & GU & 0.56  & 0.17  & 0.40  \\
   &  &                        														 MS & 0.27  & 0.33  & 0.21  \\
   \hline
\end{tabular}
\label{lagset_table}
\end{table}

\begin{table}
\caption{Effects of changing the window/block size. Empirical size and power for \citet{guan2004isotropy} [denoted GU] and \citet{maity2012test} [\citetalias{maity2012test}] for 200 realizations of a mean 0 GRF with $n = 300$ uniformly distributed sampling locations. The label ``normal" corresponds to the window/block size of $4 \times 2$, while ``small" represents using a smaller window, and ``large" denotes using a larger window.}
\centering
\begin{tabular}{| l | l | c |p{9 mm} | p{9 mm}| p{9 mm} |}
\multicolumn{6}{c}{\textbf{16 width $\times$ 10 height domain}} \\
\hline
\multicolumn{3}{|c|}{}  & \multicolumn{3}{c|}{\textbf{Window/Block Size}}  \\
\hline
$R$ & $\theta$ & Method &  small & normal  & large \\
\hline
  \parbox[t][0.75mm][c]{4.0 mm}{0} &  \parbox[t][0.75mm][c]{4.0 mm}{0} & GU & 0.06  & 0.04 & 0.01  \\
   &  &   													MS & 0.03  & 0.04  & 0.02  \\
   \hline
 \parbox[t][0.75mm][c]{4.0 mm}{$\sqrt{2}$} & \parbox[t][0.75mm][c]{4.0 mm}{$0$} & GU & 0.17  & 0.17  & 0.08  \\
   &  &                                    										  MS & 0.06 & 0.09  & 0.09  \\
   \hline
  \parbox[t][0.75mm][c]{4.0 mm}{2} & \parbox[t][0.75mm][c]{4.0 mm}{$0$} & GU & 0.56  & 0.53  & 0.22  \\
   &  &                        										MS & 0.17  & 0.17  & 0.18  \\
   \hline
\end{tabular}
\label{blocksize_table}
\end{table}

\begin{table}
\caption{Effects of changing bandwidth. Empirical size and power for \citet{guan2004isotropy} [denoted GU] for 100 realizations of a mean 0 GRF with $n = 400$ uniformly distributed sampling locations using a nominal level of $\alpha = 0.05$.}
\begin{subtable}{0.5\linewidth}
\centering
\subcaption{P-value: asymptotic $\chi^2$ distribution}
\begin{tabular}{| l | l | c |p{9 mm} | p{9 mm}| p{9 mm} |}
\multicolumn{6}{c}{\textbf{16 width $\times$ 10 height domain}} \\
\hline
\multicolumn{3}{|c|}{}  & \multicolumn{3}{c|}{\textbf{Effective Range}}  \\
\hline
$R$ & $\theta$ & Bandwidth & 3 & 6 & 12 \\
\hline
  \parbox[t][0.75mm][c]{4.0 mm}{0} &  \parbox[t][0.75mm][c]{4.0 mm}{0} & 0.65 & 0.00  & 0.00 & 0.00  \\
   &  &   															      0.75 & 0.03  & 0.06  & 0.04  \\
   &  &   															      0.85 & 0.06  & 0.11  & 0.16  \\
   \hline
\parbox[t][0.75mm][c]{4.0 mm}{$\sqrt{2}$} & \parbox[t][0.75mm][c]{4.0 mm}{$\frac{3\pi}{8}$} &  0.65 & 0.01  & 0.01  & 0.08  \\
  &  &                                    															  0.75 & 0.08 & 0.14  & 0.24  \\
   &  &                                    															0.85 & 0.14 & 0.27  & 0.35  \\
  \hline
 \parbox[t][0.75mm][c]{4.0 mm}{2} & \parbox[t][0.75mm][c]{4.0 mm}{$\frac{3\pi}{8}$} & 0.65  & 0.21  & 0.22  & 0.25  \\
  &  &                        														 0.75 & 0.50  & 0.54  & 0.67  \\
 &  &                        														 0.85 & 0.70  & 0.73  & 0.81  \\
  \hline
\end{tabular}\label{bw1}
\end{subtable}
\quad
\begin{subtable}{0.5\linewidth}
\centering
\subcaption{P-value: finite sample }
\begin{tabular}{| l | l | c |p{9 mm} | p{9 mm}| p{9 mm} |}
\multicolumn{6}{c}{\textbf{16 width $\times$ 10 height domain}} \\
\hline
\multicolumn{3}{|c|}{}  & \multicolumn{3}{c|}{\textbf{Effective Range}}  \\
\hline
$R$ & $\theta$ & Bandwidth & 3 & 6 & 12 \\
\hline
  \parbox[t][0.75mm][c]{4.0 mm}{0} &  \parbox[t][0.75mm][c]{4.0 mm}{0} & 0.65 & 0.02  & 0.03 & 0.02  \\
   &  &   															      0.75 & 0.03  & 0.06  & 0.06  \\
   &  &   															      0.85 & 0.07  & 0.10  & 0.09  \\
   \hline
\parbox[t][0.75mm][c]{4.0 mm}{$\sqrt{2}$} & \parbox[t][0.75mm][c]{4.0 mm}{$\frac{3\pi}{8}$} & 0.65 & 0.05  & 0.06  & 0.20  \\
  &  &                                    															  0.75 & 0.09 & 0.18  & 0.29  \\
   &  &                                    															 0.85 & 0.11 & 0.24  & 0.31  \\
  \hline
 \parbox[t][0.75mm][c]{4.0 mm}{2} & \parbox[t][0.75mm][c]{4.0 mm}{$\frac{3\pi}{8}$} & 0.65 & 0.37  & 0.38  & 0.53  \\
  &  &                        														 0.75 & 0.55  & 0.58  & 0.69  \\
  &  &                        														 0.85 & 0.63  & 0.64  & 0.76  \\
  \hline
\end{tabular}\label{bw2}
\end{subtable}
\label{bandwidth_table}
\end{table}

\end{document}